\documentclass[aps, floatfix, prd, twocolumn, 10pt, nofootinbib]{revtex4-1}

\usepackage{amsmath,
amssymb}
\usepackage{aas_macros}

\usepackage{bbm}
\usepackage{epsfig}
\usepackage{slashed}
\usepackage{color}
\usepackage{feynmf}
\graphicspath{{},{graphics/}}
\usepackage{hyperref}

\renewcommand{\=}{\;=\;}

\newcommand{\eq}[1]{eq.~(\ref{#1})}
\newcommand{\eqs}[1]{eqs.~(\ref{#1})}
\newcommand{\fig}[1]{fig.~\ref{#1}}

\newcommand{\Tr}{\mbox{Tr}}
\newcommand{\SNG}{\mathcal{S}}
\newcommand{\SI}{S^{-1}}
\newcommand{\sumint}{\sum\hspace{-4.5mm}\int}
\newcommand{\va}[1]{\vert\vec #1 \vert}

\newcommand{\bit}{\begin{itemize}}
\newcommand{\eit}{\end{itemize}}
\newcommand{\beq}{\begin{equation}}
\newcommand{\eeq}{\end{equation}}
\newcommand{\bea}{\begin{align}}
\newcommand{\eea}{\end{align}\]}
\newcommand{\lb}{\left(}
\newcommand{\rb}{\right)}
\def\bseq#1\eseq{\begin{equation}\begin{split}#1\end{split}\end{equation}}

\begin{document}

 \title{Dyson-Schwinger approach to color superconductivity 
at finite temperature and density}
 \date{\today}
 \author{D. M\"uller} \author{M. Buballa} \author{J. Wambach}
 \affiliation{Institut f\"ur Kernphysik (Theoriezentrum), 
Technische Universit\"at Darmstadt, Germany}

\begin{abstract}
We investigate the phases of dense QCD matter at finite temperature with 
Dyson-Schwinger equations for the quark propagator for $N_f=2+1$ flavors. 
For the gluon propagator we take a fit to quenched lattice data 
and add quark-loop effects perturbatively in a 
hard-thermal-loop--hard-dense-loop approximation. 
We consider 2SC and CFL-like pairing with chiral up and down quarks and 
massive strange quarks and present results for the condensates and the phase 
diagram. We find a dominant CFL phase at chemical potentials larger than 
$500-600$ MeV. At lower values of the chemical potential we find a 2SC phase, 
which also exists in a small band at higher temperatures for larger chemical 
potentials. With values of $20-30$ MeV, the critical temperatures to the normal phase turn out to be 
quite small.
\end{abstract}

\maketitle

\section{Introduction}
The phase diagram of strong-interaction matter (QCD matter)  is a central object in experimental and theoretical studies
 \cite{BraunMunzinger:2009zz,Fukushima:2010bq}. At low temperatures and densities, quarks and gluons are confined into hadrons while at high temperatures free quarks and gluons are the dominant degrees of freedom.

The low-density regime of QCD matter is understood quite well as it is accessible by heavy-ion collisions and, 
on the theory side, by first-principles calculations on the lattice.
These lattice QCD simulations at physical quark masses show a crossover transition from the hadronic phase at low temperatures to the quark-gluon plasma at high temperatures \cite{Bazavov:2011nk,Borsanyi:2010bp}. 

On the other hand, the regime at non-vanishing net baryon densities is much less understood. Lattice QCD fails due to the fermion sign problem that prevents Monte Carlo simulations to be performed at non-zero baryon chemical potential, and only a small window can be accessed with extrapolation methods \cite{deForcrand:2010he,Endrodi:2011gv,Kaczmarek:2011zz,Karsch:2011yq,Borsanyi:2012cr}.

The regime of large densities and low temperatures is totally inaccessible with lattice QCD. There, quarks form Cooper pairs which condense as color superconductors~\cite{Rajagopal:2000wf,Alford:2001dt,Schafer:2003vz,Rischke:2003mt,Buballa:2003qv,Shovkovy:2004me,Alford:2007xm}.
At very high densities the coupling becomes small and QCD can be studied 
with weak-coupling methods. In this way, it was shown that the ground state 
is a color superconductor in the color-flavor-locked (CFL) phase, where up,
down, and strange quarks are paired 
symmetrically~\cite{Alford:1998mk,Schafer:1999fe,Shovkovy:1999mr}. 
On the other hand,
at quark chemical potentials below 1~GeV, as relevant for
the description of compact-star interiors, weak-coupling methods are not 
applicable, and one has to resort to non-perturbative methods.
In this regime the strange-quark mass cannot be neglected 
against the chemical potential, eventually making the CFL pairing pattern
unfavorable. This has been studied extensively in effective models,
like the Nambu--Jona-Lasinio (NJL) model. It was found that at intermediate chemical potentials the
2SC phase is favored~\cite{Buballa:2001gj}, in which only up and down quarks
form Cooper pairs.
Moreover, when the constraints of beta equilibrium and
electric neutrality are taken into account in addition, these model studies show a rich 
phase structure with many different types of color-superconducting 
condensates \cite{Abuki:2005ms,Ruester:2006aj,Blaschke:2005uj}. 

Unfortunately, these results are very sensitive to the model parameters.
In order to make quantitative predictions, it is therefore necessary to 
employ non-perturbative methods which are directly rooted in QCD.
In this context, Dyson-Schwin\-ger equations (DSEs) are a promising approach. 
They have been used to investigate the chiral and deconfinement transition 
and good agreement with lattice results was 
obtained~\cite{Fischer:2009gk,Fischer:2010fx}. 
In Refs.~\cite{Fischer:2011mz,Fischer:2012vc},
these studies were extended to finite density.
It was found that the resulting phase diagram contains a 
critical endpoint beyond which the chiral phase transition becomes
first-order.

Some time ago, the framework of DSEs was also applied to investigate 
color-superconducting phases at zero 
temperature~\cite{Nickel:2006vf,Nickel:2006kc,Nickel:2008ef}.
In contrast to the NJL-model studies, these calculations showed that 
the CFL phase is the dominant phase down to rather low chemical potentials,
leaving practically no room for other quark phases, like the 2SC phase. 

In their exact formulation, DSEs are an infinite set of equations, 
which have to be truncated in practice.
This should be done in a way that the most important features of QCD are
included, while keeping the numerical effort on a reasonable level.
Obviously, the truncated DSEs are no longer exact and typically require
some model input for higher-order n-point functions.  
However, unlike effective models, they are applicable at all energy scales
and the approximations are systematically amendable.

Recently, considerable progress was made in extracting some of
the relevant input from the lattice. 
Specifically, there exists a temperature dependent parametrization of
the quenched gluon propagator~\cite{Fischer:2010fx}, which,
together with a phenomenological ansatz for the quark-gluon vertex,
has successfully been used as input for the DSE calculations of 
Ref.~\cite{Fischer:2011mz,Fischer:2012vc}.
Compared with earlier truncations, these gluon interactions turn out
to be more attractive, leading to stronger non-perturbative effects.

One goal of the present paper is therefore to perform an update of
the DSE analysis of color-superconducting phases presented in 
Ref.~\cite{Nickel:2006kc}, using this new, more realistic gluonic input.
We will show that this changes the results qualitatively,
making the existence of a 2SC phase at low temperatures possible.
The second goal of this work is to extend these calculations to non-zero 
temperature, which was not studied in Ref.~\cite{Nickel:2006kc}.

This paper is organized as follows.
In  sects.~\ref{sec_dse} and \ref{sec_struc}, we introduce the 
quark DSEs and the structure of the quark propagators in a color 
superconducting environment.
Our truncation scheme is presented in sect.~\ref{sec_hdl}. 
In sect.~\ref{sec_results}, we discuss our results for the
color-super\-conducting condensates and for the phase diagram.
All calculations are performed for $N_f=2+1$ flavors.
Thereby, we restrict ourselves to the case of a flavor-independent
chemical potential, postponing the analysis of electrically neutral
matter to a later analysis. 
In sect.~\ref{sec_vacparam} we discuss the phase diagram obtained with 
an alternative parametrization of the quark-gluon vertex and the gluon propagator, 
which leads to a better fit of vacuum quantities.
Conclusion are drawn in sect.~\ref{sec_conclusions}.

\section{The quark Dyson-Schwinger equation}\label{sec_dse}

\begin{figure}
	\centering
		\includegraphics[scale=0.9]{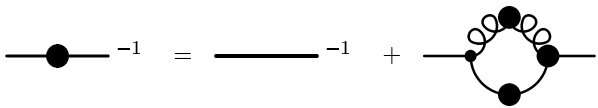}	
\caption{Dyson-Schwinger Equation for the full quark propagator. Plain lines represent quark propagators, the curly line the gluon propagator. Thick dots represent dressed quantities.}
\label{feyn_qdse}
\end{figure}

The dressed quark propagator $\SNG(p)$ is the solution of the Dyson-Schwinger 
equation (DSE) 
\beq\label{GE}
\SNG^{-1}(p) \= Z_2(\SNG^{-1}_0(p) + \Sigma(p)),
\eeq
which is depicted diagrammatically in \fig{feyn_qdse}.
Here $p=(p_4,\vec p)$ is the Euclidean 4-momentum. 
To describe the system 
at temperature $T$ and chemical potential $\mu$ we use the Matsubara formalism,
so that $p_4 = \omega_n + i\mu$ where 
$\omega_n=(2n+1)\pi T$ are the fermionic Matsubara frequencies.
$\SNG_0(p)$ denotes the bare quark propagator and $Z_2$ is the quark wave 
function renormalization constant.

The self-energy $\Sigma(p)$ is given by
\beq\label{SE}
Z_2\Sigma(p) \= g^2\sumint\limits_q\,\Gamma_\mu^{a,0}\SNG(q)\Gamma^b_\nu(p,q)D^{ab}_{\mu\nu}(k=p-q)
\eeq
with the dressed gluon propagator $D^{ab}_{\mu\nu}(k)$
and the bare and dressed quark-gluon vertices, 
$g\Gamma^{a,0}_\nu$ and $g\Gamma^b_\nu(p,q)$, respectively. 
$a$ and $b$ denote color indices, $\mu$ and $\nu$ Dirac indices
and $g$ represents the QCD coupling constant. 

Furthermore, we have introduced the notation
${\sum\hspace{-3.5mm}\int\hspace{1mm}}_q \equiv T\sum\int\frac{d^3q}{(2\pi)^3}$,
where the sum runs over the Matsubara frequencies corresponding to $q_4$.

As we work in Landau gauge, the gluon propagator is transverse. 
In a thermal medium it can have two components in the transverse subspace,
\beq
\label{eq:Dfull}
D^{ab}_{\mu\nu}(k) \= \frac{Z^{ab}_{TT}(k)}{k^2} P^T_{\mu\nu}(k) + \frac{Z^{ab}_{TL}(k)}{k^2} P^L_{\mu\nu}(k),
\eeq
with the 3-dimensional transverse and longitudinal projectors 
\bseq
P^T_{\mu\nu}(k) &\= \delta_{ij}-\frac{k_i k_j}{\vec k^2},\\
P^L_{\mu\nu}(k) &\= T_{\mu\nu}(k) - P^T_{\mu\nu}(k)
\eseq
and the 4-dimensional transverse projector
\beq
T_{\mu\nu}(k) \= \delta_{\mu\nu}-\frac{k_\mu k_\nu}{k^2}.
\eeq
The gluon dressing functions $Z^{ab}_{TT}(k)$ and $Z^{ab}_{TL}(k)$ coincide 
in vacuum but, in general, differ in the medium, 
as the medium breaks the $O(4)$ symmetry of the vacuum.

The quark spinors are four-dimensional objects in Dirac space and have
 $3\times N_f$ components in color-flavor space,
where $N_f$ is the number of flavors.
In this article we generally consider $N_f = 3$,
but sometimes refer to two-flavor systems for comparison.
In addition we introduce the two-dimensional Nambu-Gorkov (NG) space, 
which easily allows to incorporate color 
superconductivity in the formalism \cite{Nambu:1960tm,Gorkov}.
The NG spinors are defined by

\beq\label{ngs}
  \Psi \= \frac{1}{\sqrt{2}}\begin{pmatrix} \psi \\ C\bar\psi^T \end{pmatrix}, ~~~\bar\Psi \= \frac{1}{\sqrt{2}}\begin{pmatrix} \bar\psi & \psi^T C \end{pmatrix}
\eeq
with the charge conjugation matrix $C=\gamma_2\gamma_4$.

We take the conventions from \cite{Rischke:2003mt,Nickel:2006vf} and 
parametrize the propagators and self-energies in NG space as follows:
\beq
 \SNG_0(p) \= \begin{pmatrix} S_0^+(p) & 0\\ 0 & S_0^-(p) \end{pmatrix}
\eeq
\beq
  \SNG(p) \= \begin{pmatrix} S^+(p) & T^-(p) \\ T^+(p)  & S^-(p) \end{pmatrix}
\eeq
\beq
   \Sigma(p) \= \begin{pmatrix} \Sigma^+(p) & \Phi^-(p) \\ \Phi^+(p)  & \Sigma^-(p) \end{pmatrix}
\eeq

The components diagonal in NG space represent the normal propagators and 
self-energies for particles ($+$) and charge conjugate particles ($-$),  
while the off-diagonal components are related to color superconductivity. 
Therefore the bare propagator is diagonal in NG space. 
Without color superconductivity, also the dressed propagator and the
self-energy are diagonal, and the Dyson-Schwinger system decouples into two 
equivalent gap equations for the quark propagator and the charge conjugate 
propagator. On the other hand, if color-superconducting condensates are 
present, quarks and charge conjugate quarks are coupled. 

This becomes evident when we insert the above expressions into the DSE, \eq{GE}. One then obtains the following set of equations,
which are coupled through the color-superconducting condensates:
\bseq
S^{\pm -1} &\= Z_2\left( S_0^{\pm -1} + \Sigma^\pm - \Phi^\mp\left(S_0^{\mp -1} + \Sigma^\mp\right)^{-1}\Phi^\pm \right) \\
T^\pm &\= - \left( S_0^{\mp -1} + \Sigma^\mp\right)^{-1} \Phi^\pm S^\pm
\label{dseng}
\eseq

By construction, the two NG-spinors \eq{ngs} are not independent.
As a consequence, the $+$ and $-$ components of the propagators
are related to each other by
\bseq
 S^+(p) &= -C S^-(-p)^T C,\\
 T^+(p) &= -C T^+(-p)^T C,
\eseq
\bseq
 S^+(p_4, \vec p) &= \gamma_4 S^+(-p_4, \vec p)^\dagger \gamma_4,\\
 T^+(p_4, \vec p) &= \gamma_4 T^-(-p_4, \vec p)^\dagger \gamma_4,
\eseq
and analogously for the self-energies~\cite{Rischke:2003mt,Nickel:2006vf}.

The bare quark propagator is defined by
\beq
(S_0^+)^{-1}(p) \= -i\slashed p + Z_{m} m_f
\eeq
with the bare quark mass $m_f$ and the mass 
renormalization constant $Z_{m}$ for the quark flavor $f$.
It follows that $(S_0^-)^{-1}$ takes the same form with $\mu$ replaced by 
$-\mu$. 

The quark renormalization constants $Z_2$ and $Z_{m}$ are determined by 
the renormalization condition in vacuum
\beq
\SI(p^2=\nu^2) \= (-i\slashed p + m_f)\vert_{p^2=\nu^2}
\eeq
at some arbitrary renormalization point $\nu$. 

The vertices also live in NG space. The bare vertex is diagonal
\beq
\label{eq:Gamma0}
\Gamma^{a,0}_{\mu} \= Z_{1F}
\begin{pmatrix} \gamma_\mu\frac{\lambda_a}{2} & 0
\\ 
0 & -\gamma_\mu\frac{\lambda_a^{T}}{2} \end{pmatrix} 
=: Z_{1F}\gamma_\mu\frac{\Lambda_a}{2},
\eeq
where $\lambda_a$ are the 8 Gell-Mann matrices in color space.
$Z_{1F}$ is the renormalization constant of the quark-gluon vertex. 
The full vertex has off-diagonal elements in general,
\beq\label{vertex_full}
  \Gamma^a_{\mu}(p,q) \= \begin{pmatrix} \Gamma^{a,+}_\mu(p,q) & \Delta^{a,-}_\mu(p,q)\\ \Delta^{a,+}_\mu(p,q) &  \Gamma^{a,-}_\mu(p,q) \end{pmatrix}.
\eeq
In this work, we only take into account the diagonal functions $\Gamma^\pm$
which will be specified later, when we define our truncation scheme.


\section{Parametrization of color-superconducting phases}\label{sec_struc}
In the following we restrict our calculations to spin-zero color-superconducting 
phases, which should be preferred over spin-one phases in most regions of the 
phase diagram. 
We are interested in the realistic case with two light quark flavors 
and a heavier strange quark. 
For simplicity, we consider massless up and down quarks but massive
strange quarks. This corresponds to an intermediate situation between
two idealized cases:

For infinitely heavy strange quarks, the strange sector decouples,
and we are left with a two-flavor system. 
In this case, the 2SC phase is favored, where 
red and green up and down quarks form a diquark condensate,
whereas blue quarks are not involved in the pairing.
This leads to a breaking of the $SU_c(3)$ color symmetry to a $SU_c(2)$ 
subgroup, while the $SU_L(2) \otimes SU_R(2)$ chiral symmetry remains intact.
The other extreme is to have three massless quark flavors. 
Here the favored phase is the CFL phase \cite{Alford:1998mk},
where all colors and flavors participate in the pairing in a symmetric way, 
so that the $SU_c(3)$ color symmetry and the $SU_L(3) \otimes SU_R(3)$ 
chiral symmetry are broken down to a residual $SU_{c+V}(3)$ symmetry.

Turning on a finite but small strange-quark mass, a CFL-like phase
can still be realized. In this phase all colors and flavors are
paired in a diquark condensate, but there is only a residual
$SU_{c+V}(2) \otimes U_{c+V}(1)$ symmetry, while the full
CFL symmetry is only approximate.
Alternatively there could be a 2SC phase, where the strange quarks
remain unpaired. 

In order to find the solutions of the DSE in a given
color-superconducting phase,
we follow Ref.~\cite{Nickel:2006kc} and expand the
propagators and self-energies as
\bseq
\label{eq:SPTM}
{S^+}(p)&
\= \sum_i {S^+_i}(p)\,P_i,
\\
T^+(p) &
\= \sum_i T^+_i(p)\,M_i,
\eseq
and
\bseq
\label{eq:SigmaPPhiM}
\Sigma^+(p) &
\=  \sum_i \Sigma^+_i(p)\,P_i,
\\
\Phi^+(p) &
\=  \sum_i \phi^+_i(p)\,M_i ,
\eseq
with $P_i$ and $M_i$ being matrices color-flavor space.
These matrices have been constructed such that they respect the symmetries 
of the considered phase and are complete in the sense that they yield 
a closed set of self-consistency equations when inserted into \eq{dseng}.
In the most general case, considered here, \textit{i.e.}, the CFL-like phase
with non-vanishing strange-quark mass, both sets, $\{P_i\}$ and $\{M_i\}$,
consist of seven matrices, which are listed in Appendix \ref{cfmatrices}.
More symmetric cases, such as the 2SC phase or non-superconducting phases,
are contained in this parametrization as special limits, which are 
discussed in Appendix \ref{cfmatrices} as well.

The Dirac structure of the propagator is parametrized as \cite{Pisarski:1999av,Nickel:2006kc}

\bseq\label{prop_dirac}
&{S_i^+}^{-1}(p) \= \\& -i(\omega_n+i\mu)\gamma_4 C_i^+(p) -i\slashed{\vec p}A_i^+(p)+B_i^+(p) -i \gamma_4\frac{\slashed{\vec p}}{\va p} D_i^+(p),\\
&T_i^+(p) \= \\&\left(\gamma_4\frac{\slashed{\vec p}}{\va p}T_{A,i}^+(p) + \gamma_4T_{B,i}^+(p) + T_{C,i}^+(p)+\frac{\slashed{\vec p}}{\va p}T_{D,i}^+(p)\right)\gamma_5 .
\eseq

$B_i^+(p)$ accounts for chiral symmetry breaking in the normal propagator. 
$D_i^+(p)$ as well as $T_{B,i}^+(p)$ and $T_{D,i}^+(p)$ are only non-zero for 
color-superconducting phases with finite strange-quark masses. 
The self-energies are decomposed analogously.

The renormalization-point dependent light-quark condensate in the chiral limit is defined by
\beq\label{eq:qq_cond}
\langle \bar u u \rangle = \langle \bar d d \rangle 
= -Z_m Z_2 \sumint\limits_q \Tr_{D,c}(S_{up}^+(q)).
\eeq
where the Dirac and color trace of the up quark component of the quark propagator is performed.

Diquark condensates can be calculated analogously as
\beq\label{condensate}
{\cal C}_i
\equiv
\langle \psi^T C\gamma_5 \mathcal{O}_i \psi \rangle 
= -Z_2 \sumint\limits_q \Tr(\gamma_5\mathcal{O}_i T^-(q))
\eeq
where $\mathcal{O}_i$ is an operator in color-flavor space.
In particular, we define ${\cal C}_{ud}$ with
\beq
    \mathcal{O}_{ud} = \frac{\tau_2}{2} \otimes \frac{\lambda_2}{2}
    = \frac{1}{4}(M_1 - M_2)\,,
\eeq
describing the mutual pairing of (red and green) non-strange quarks
in the 2SC phase as well as in the CFL phase,
and ${\cal C}_{uds}$ with
\bseq
    \mathcal{O}_{uds} &= \frac{1}{2}\left(
                        \frac{\tau_5}{2} \otimes \frac{\lambda_5}{2}
                        +
                        \frac{\tau_7}{2} \otimes \frac{\lambda_7}{2}
                        \right)
\\
    &= \frac{1}{8}(M_6 + M_7 - M_4 -M_5)\,,
\label{eq:cond_uds}
\eseq
describing the pairing of non-strange quarks with strange quarks
in the CFL phase.

These condensates represent the dominant pairing patterns in the
color-superconducting phases under consideration.
In a phase with exact CFL symmetry, 
$\mathcal{C}_{ud}$ and $\mathcal{C}_{uds}$ are equal, 
while in a CFL-like phase with heavier strange quarks, they can be different.
In the 2SC phase, finally, $\mathcal{C}_{uds}$ vanishes.

\section{Truncation scheme}
\label{sec_hdl}

The quark DSE (\ref{GE}) and all other equations discussed so far 
are exact QCD equations.
However, in order to solve them, we need to specify the dressed gluon 
propagator and the dressed quark-gluon vertex, which enter the 
self-energy, \eq{SE}.
In principle, they are given by their own DSEs
but, since these involve even higher n-point functions, 
an exact solution would result in an infinite tower of equations.
Therefore we have to rely on truncations.
In this section we specify the truncation scheme used in the present work.

\subsection{Dressed quark-gluon vertex}
\label{sec_vertex}

We begin with the dressed quark-gluon vertex, \eq{vertex_full}.
We restrict it to the diagonal NG components and an abelian ansatz 
\beq
\Gamma^a_\mu(p,q) \= \frac{\Lambda_a}{2}\gamma_\mu \Gamma(k),
\eeq
with $\Lambda_a$ as defined in \eq{eq:Gamma0}
and a dressing function $\Gamma$,
which only depends on the gluon momentum $k = p-q$.
For this function, we take a model ansatz, 
similar to the vertex proposed in Ref.~\cite{Fischer:2009gk}:
\beq\label{vertex1}
\frac{\Gamma(k)}{Z_2\tilde Z_3} = \frac{d_1}{d_2+k^2} + \frac{k^2}{k^2+\Lambda^2}\left(\frac{\beta_0\alpha(\nu)\ln(k^2/\Lambda^2+1)}{4\pi}\right)^{2\delta}.
\eeq
Here $\tilde Z_3$ is the ghost propagator renormalization constant.
The second term in parentheses describes the perturbative running in the 
ultraviolet, with $\beta_0 = (11 N_c-2N_f)/3$, $\delta = -9N_c/(44N_c-8N_f)$ and a scale factor
$\Lambda$. $\alpha(\nu) = g^2/4\pi$ is the strong fine-structure constant
at the renormalization scale $\nu$.
Following Ref.~\cite{Fischer:2009gk}, we take $\alpha(\nu) = 0.3$ and
$\Lambda = 1.4$~GeV. 

For the first term, which is relevant for the infrared behavior,
we take $d_2 = 0.5$ GeV$^2$~\cite{Fischer:2009gk} and fit $d_1$ to obtain
a critical temperature of around $150$ MeV for the chiral phase transition
at $\mu = 0$. We get a value of $d_1=9.6$ GeV$^2$. 
As we will show in sect.~\ref{sec_vacparam} this parametrization yields
too large values for the chiral condensate and the pion decay constant in vacuum,
whereas a fit to these quantities would yield a too low critical temperature.
Since the focus of this article lies on the phase diagram, we consider a realistic critical
temperature to be more important. We therefore continue with the parameters specified above. 
However, we will come back to this issue in sect.~\ref{sec_vacparam}, where we
introduce an alternative parametrization, fitted to vacuum quantities. 

The vertex in Ref.~\cite{Fischer:2009gk}
has additional contributions motivated by a Slavnov-Tay\-lor identity and
a Ball-Chiu vertex construction \cite{Ball:1980ay}. As there is no strict argument
for the necessity of these components, we dropped them as they lead to instabilities
in the iteration of color-super\-conducting phases.

In Landau gauge, the renormalization constant $Z_{1F}$, which appears 
in the bare vertex \eq{eq:Gamma0},
can be expressed as $Z_{1F} = Z_2/\tilde Z_3$ 
using Slavnov-Taylor identities.
The normal and anomalous components of the self-energy \eq{SE} 
then become
\bseq
\Sigma^+(p) &\= 
\phantom{-} 4\pi \alpha(\nu) \sumint\limits_q \frac{\Gamma(k)}{\tilde Z_3}
\gamma_\mu\frac{\lambda_a}{2}S^+(q)\gamma_\nu\frac{\lambda_b}{2}
D^{ab}_{\mu\nu}(k),
\\
\Phi^+(p) &\= 
- 4\pi \alpha(\nu) \sumint\limits_q \frac{\Gamma(k)}{\tilde Z_3}
\gamma_\mu\frac{\lambda_a^T}{2}T^+(q)\gamma_\nu\frac{\lambda_b}{2}
D^{ab}_{\mu\nu}(k),
\eseq
where $k$ is again the gluon momentum $p-q$.
Note that $\tilde Z_3$ drops out when \eq{vertex1} is inserted.

\subsection{Dressed gluon propagator}

\begin{figure}[t!]
	\centering
		\includegraphics[scale=0.9]{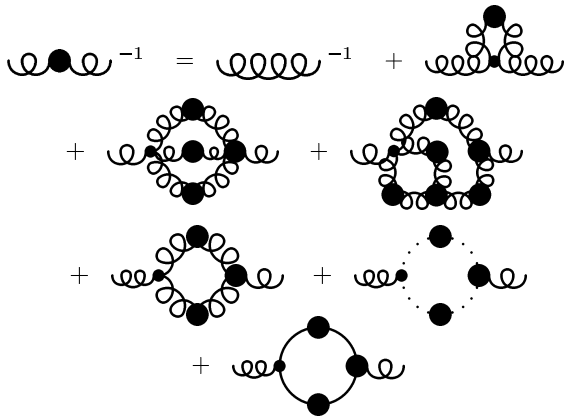}	
\caption{DSE for the gluon propagator. Curly, dotted and plain lines represent gluon, 
ghost, and quark propagators, respectively. Thick dots indicate dressed quantities.}
\label{feyn_gdse_full}
\end{figure}

\begin{figure}
	\centering
		\includegraphics[scale=0.9]{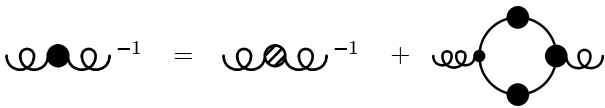}	
\caption{Unquenching of the gluon propagator. The shaded propagator is the full Yang-Mills gluon propagator.}
\label{feyn_gdse}
\end{figure}

The gluon DSE is depicted in \fig{feyn_gdse_full}. 
Restricting this equation to the first three lines corresponds 
to the pure (``quen\-ched'') Yang-Mills system, 
while the last diagram describes the coupling to the quarks.
Solving the Yang-Mills equations is numerically demanding already without 
quarks~\cite{Maas:2004se,Maas:2005hs,Cucchieri:2007ta}.
In recent years significant progress has been made in this sector by combining
continuum methods with lattice calculations. 
Therefore we consider a truncated version of the full gluon DSE,
where we take the solution of the quenched Yang-Mills equations as input
and include the quark effects only perturbatively. 
This scheme is depicted in \fig{feyn_gdse}.
It neglects back-coupling effects of the quark propagator on the Yang-Mills 
sector but should carry the dominant effects of the quark loop on the 
gluon propagator.

The gluon DSE is then given by
\beq
\label{gdse}
  D^{-1,ab}_{\mu\nu}(k) = D^{-1,ab}_{\mu\nu, YM}(k) + \Pi^{ab}_{\mu\nu}(k)
\eeq
with the Yang-Mills gluon propagator  $ D^{ab}_{\mu\nu, YM}$ and the 
gluon polarization tensor $\Pi^{ab}_{\mu\nu}$. 
The former has the same transverse Dirac structure as the full gluon 
propagator, \eq{eq:Dfull}, and is diagonal in color space,
\beq
\label{eq:DYM}
  D^{ab}_{\mu\nu,YM}(k) = \lb\frac{Z_{TT}^{YM}(k)}{k^2} P^T_{\mu\nu}(k) 
  +\frac{Z_{TL}^{YM}(k)}{k^2} P^L_{\mu\nu}(k)\rb\delta^{ab}.
\eeq
For the dressing functions we adopt the parametrizations of 
Ref.~\cite{Fischer:2010fx},
\bseq\label{eq:ZYM}
Z_{TT,TL}^{YM}(k) &\= \frac{k^2\Lambda^2}{(k^2+\Lambda^2)^2}\Bigg(\left(\frac{c}{k^2+a_{T,L}\Lambda^2}\right)^{b_{T,L}} \\&\;+\; \frac{k^2}{\Lambda^2}\left(\frac{\beta_0\alpha(\nu)\ln(k^2/\Lambda^2+1)}{4\pi}\right)^\gamma\Bigg),
\eseq
which have been obtained by fitting lattice results.
The scale factor $\Lambda$ and the parameters describing the perturbative
behavior in the UV are the same as in \eq{vertex1}. The exponent 
$\gamma = (-13N_c+4N_f)/(22N_c-4N_f)$ is related to $\delta$ by $2\delta + \gamma = -1$.
The infrared behavior is parametrized by a temperature-independent
constant $c= 11.5$~GeV$^2$ and temperature-dependent parameters
$a_{T,L}$ and $b_{T,L}$, which are tabulated in Ref.~\cite{Fischer:2010fx}
for various temperatures. We determine their values at other temperatures by 
linear interpolation.

The polarization tensor is given by 
\beq
  \label{polarization}
  \Pi^{ab}_{\mu\nu}(k) \= -\frac{g^2}{2}\sumint\limits_q\Tr \left(\Gamma_{\mu}^{a,0}\SNG(p)\Gamma_{\nu}^{b}(p,q)\SNG(q) \right)
\eeq
with $p = k + q$. 
It depends, in principle, on the dressed quark propagator, but, as a further simplification, 
we will evaluate it with bare quark propagators in HTL-HDL approximation (see sect.~\ref{sec:htl}).
The trace has to be carried out in Dirac, color, flavor and Nambu-Gorkov 
space.
As quantum corrections must not change the transverse nature of the gluon, 
see Eqs.~(\ref{eq:Dfull}) and (\ref{eq:DYM}), 
we require the polarization tensor to be transverse as well, \textit{i.e.},
\beq
\Pi^{ab}_{\mu\nu}(k) \= 
\Pi^{ab}_{TT}(k) P^T_{\mu\nu}(k) + \Pi^{ab}_{TL}(k) P^L_{\mu\nu}(k).
\eeq
The dressed gluon propagator is then given by
\bseq
\label{eq:Ddressed}
  D^{ab}_{\mu\nu}(k) &\= \frac{Z_{TT}^{YM}(k)}{k^2 + Z_{TT}^{YM}(k)\Pi^{ab}_{TT}(k)} P^T_{\mu\nu}(k) \\&\;+\;\frac{Z_{TL}^{YM}(k)}{k^2 + Z_{TL}^{YM}(k)\Pi^{ab}_{TL}(k)} P^L_{\mu\nu}(k).
\eseq

\subsection{HTL-HDL approximation}\label{sec:htl}

Since the gluon polarization tensor \eq{polarization} depends on the dressed quark 
propagator, \eq{eq:Ddressed} must, in principle, 
be solved self-consistently together with the quark DSE. 
However, in a first step, we perform a simpler non-self-consistent
approximation, which was also employed in Ref.~\cite{Nickel:2006vf,Nickel:2006kc}.

In this scheme, we neglect the vacuum part of the polarization loop
and calculate the medium corrections in hard-thermal-loop--hard-dense-loop
(HTL-HDL) approximation, using the bare propagator of massless quarks,
\beq\label{Sbare}
(S^+)^{-1}(p) = -i(\omega_n+i\mu)\gamma_4 - i\slashed{\vec p}.
\eeq
In the HTL-HDL approximation, it is assumed that the external momenta are 
small compared to the chemical potential or temperature.
This allows for an approximate analytical evaluation of the loop integral.
The result can be found in textbooks, \textit{e.g.}, Ref.~\cite{Bellac:2000b}, 
and is given by 
\bseq\label{hdl}
&Z_{TT}^{YM}(k)\Pi^{ab}_{T}(k) \=\\ &~~~ m_{TT}^2\frac{\omega_m}{\va k}\left[\left(1+\left(\frac{\omega_m^2}{\vec k^2}\right)\right)iQ\left(\frac{i\omega_m}{\va k}\right) -\frac{\omega_m}{\va k}\right]\delta^{ab}\\
&Z_{TL}^{YM}(k)\Pi^{ab}_{L}(k) \=\\ &~~~ 2m_{TL}^2\frac{\omega_m^2+{\vec k}^2}{{\vec k}^2}\left[1-\frac{\omega_m}{\va k}iQ\left(\frac{i\omega_m}{\va k}\right)\right]\delta^{ab}
\eseq
with
\beq
iQ(ix) \= \frac{i}{2}\ln\frac{ix+1}{ix-1} = \arctan\left(\frac{1}{x}\right).
\eeq
and bosonic Matsubara frequencies $\omega_m=2m\pi T$.
The effective transverse and longitudinal gluon masses are defined by
\beq\label{mhdl}
m_{TT,TL}^2 \= N_f\alpha_{TT,TL}(k)\left(\frac{\pi T^2}{3} + \frac{\mu^2}{\pi}\right)
\eeq
with the renormalization-point independent running couplings, 
which are given by~\cite{Nickel:2006vf} 
\beq
\alpha_{TT,TL}(k) \= 
\frac{\Gamma(k)Z^{YM}_{TT,TL}(k)}{Z_2\tilde Z_3}\alpha(\nu).
\eeq

In this approximation, all quark-mass effects and color-superconducting 
contributions to the polarization tensor are neglected,
but, as it provides analytical results for the gluon polarization,
it has the advantage to keep the numerical effort at the same level 
as in a pure rainbow truncation  with quenched gluons.

An extended truncation scheme, where the solutions of the quark DSE
are self-consistently used in the gluon polarization tensor,
will be discussed in a forthcoming paper~\cite{Mueller:2013ip}.

\subsection{Strange-quark mass}

The last input quantity to be specified is the strange-quark mass at the 
renormalization scale $\nu$. 
According to the particle data group (PDG) \cite{Beringer:1900zz}, 
its value is $m_s=95\pm 5$ MeV in the $\overline{MS}$ renormalization scheme 
at a renormalization scale of $\nu=2$ GeV. 
In earlier Dyson-Schwinger calculations \cite{Nickel:2006kc} the 
renormalization point was therefore chosen to be $\nu=2$~GeV as well,
and the PDG value for $m_s$ was directly used
as an input.

\begin{figure}
	\centering
	\includegraphics[]{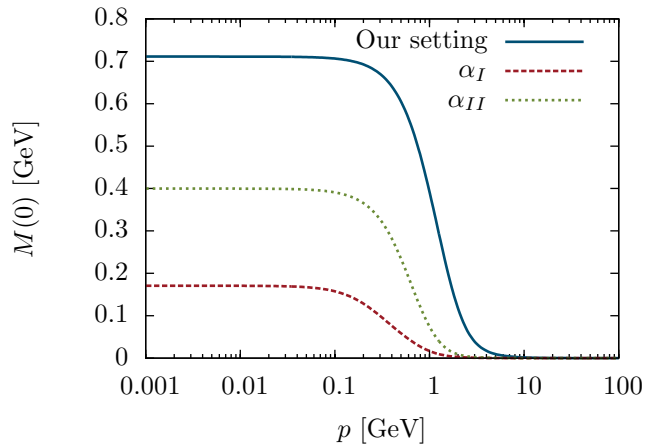}
	\caption{Vacuum-mass functions $M(p) = B_1^+(p)/A_1^+(p)$ 
                 of chiral quarks ($m=0$) for the interactions
                 $\alpha_{I,II}$ from \cite{Nickel:2006kc} and in
                 our setting.}
\label{Mvac_comp}
\end{figure}

However, as already pointed out in the Introduction,
the gluon dressing functions and vertices used in 
Ref.~\cite{Nickel:2006kc} differ from ours, which has a strong
effect on the mass functions. 
This is illustrated in \fig{Mvac_comp}, where the vacuum-mass functions 
of the (chiral) up and down quarks are shown for the two interactions
used in Ref.~\cite{Nickel:2006kc} and for the present setting.

An essential difference is that, for the former, the 
chiral symmetry restoration sets in at lower momenta.
As a consequence, $p=2$~GeV already belongs to the
perturbative regime, so that the perturbative strange-quark mass could be used at this scale.
In contrast, with the improved parametrizations of quark-gluon vertex and 
gluon propagator used in our calculations,
there are still considerable non-perturbative effects at $2$~GeV, 
and therefore we have to renormalize at a higher scale to 
be in the perturbative region. To be on the safe side, we choose 
$\nu=100$ GeV. 

The renormalization scheme used in this work is a momentum subtraction (MOM) scheme.
We can therefore take the MOM-scheme value of the strange-quark mass obtained 
in the lattice calculation of Ref.~\cite{Durr:2010vn} and evolve it to our renormalization point 
$\nu=100$~GeV with a four-loop running \cite{Chetyrkin:1999pq}. Assuming an error band 
of the same order as in the PDG, we find a strange-quark mass $m_s(100\mbox{~GeV})=55-62$~MeV.

On the other hand, the scale factor $\Lambda=1.4$~GeV in Eqs.~(\ref{vertex1}) and (\ref{eq:ZYM}),
which was taken from Ref.~\cite{Fischer:2009gk}, is higher than the expected MOM scale,
and it is not clear whether the perturbatively evolved strange-quark mass is consistent with our truncation. 
A larger scale leads to a stronger increase of the quark mass in the infrared. 
This could be important, as a large strange-quark mass disfavors the  CFL phase relative 
to the 2SC phase. 
Since, as mentioned in the Introduction, the existence of the 2SC phase at intermediate 
chemical potentials will be our central result, we have to make sure that this is not an
artifact of a too large choice for $m_s$. 
We therefore take the above result of the evolution as an upper limit and additionally consider
a smaller strange-quark mass to check the stability of the 2SC phase.  
To be specific, we perform calculations with $m_s = 54$~MeV and with $m_s = 30$~MeV
at $\nu = 100$~GeV.

\begin{figure}
    \centering
    	\includegraphics[]{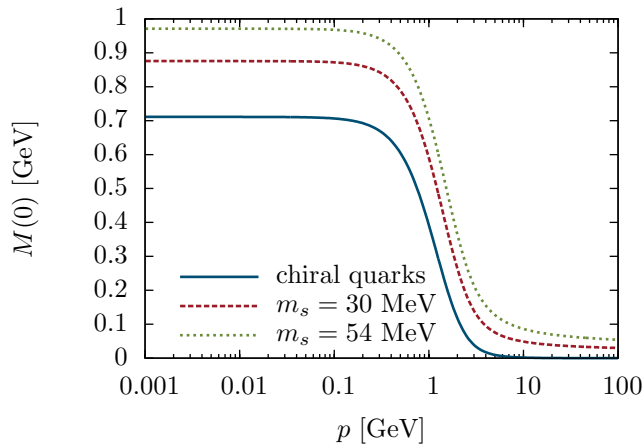}
    \caption{Vacuum-mass functions $M(p)$
                 of chiral quarks ($m=0$) and strange quarks ($m=30$ MeV, $m=54$ MeV).}
\label{Mvac_s}
\end{figure}

The corresponding mass functions $M(p)$ in vacuum are shown in \fig{Mvac_s}, 
together with the mass function of the chiral ($m=0$) up and down quarks.
For the latter we find $M(0)=710$~MeV, while for the strange quarks we obtain
$M(0)=875$~MeV for $m_s = 30$~MeV and $M(0)=970$~MeV for $m_s = 54$~MeV.
These values seem to be unrealistically large, when we compare them
with typical constituent quark masses in phenomenological models. 
Indeed, as mentioned earlier, our fit of the vertex parameters to the chiral critical 
temperature at $\mu = 0$ tends to overestimate the strength of chiral symmetry breaking 
in vacuum (see sect.~\ref{sec_vacparam}). 
In an improved truncation scheme, which will be discussed in a future 
publication~\cite{Mueller:2013ip}, we obtain smaller masses at $p=0$, 
which are of the order of $450$~MeV for up and down quarks 
and $600$ to $700$~MeV for strange quarks. 
It should be emphasized, however, that the quark masses are not observable.
Moreover, the size of the non-perturbative regime will roughly stay the same.

\section{Results}\label{sec_results}
\begin{figure}
	\centering
	    \includegraphics[]{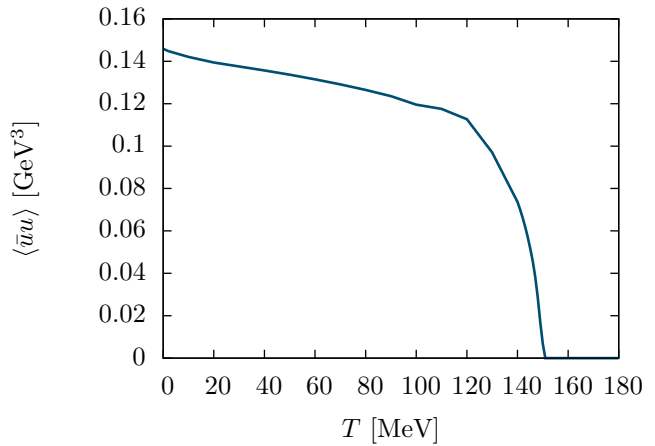}
	\caption{Temperature-dependence of the chiral condensate $\langle \bar u u \rangle$ at the renormalization scale $\nu=100$~GeV at $\mu=0$.}
\label{fig:condT}
\end{figure}

The temperature dependence of the (non-strange) chiral condensate 
at $\mu = 0$ is shown in \fig{fig:condT}.
The condensate is decreases with increasing
temperature until chiral symmetry is restored in a second-order phase transition at around $T=150$~MeV. The small kinks are due to the temperature-dependent lattice input.

\begin{figure}
	\centering
	    \includegraphics[]{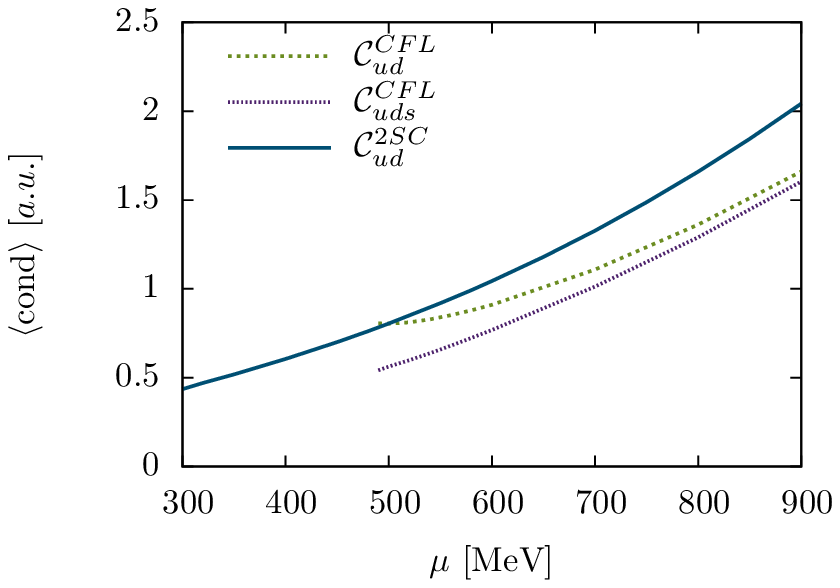}
	\centering
		\includegraphics[]{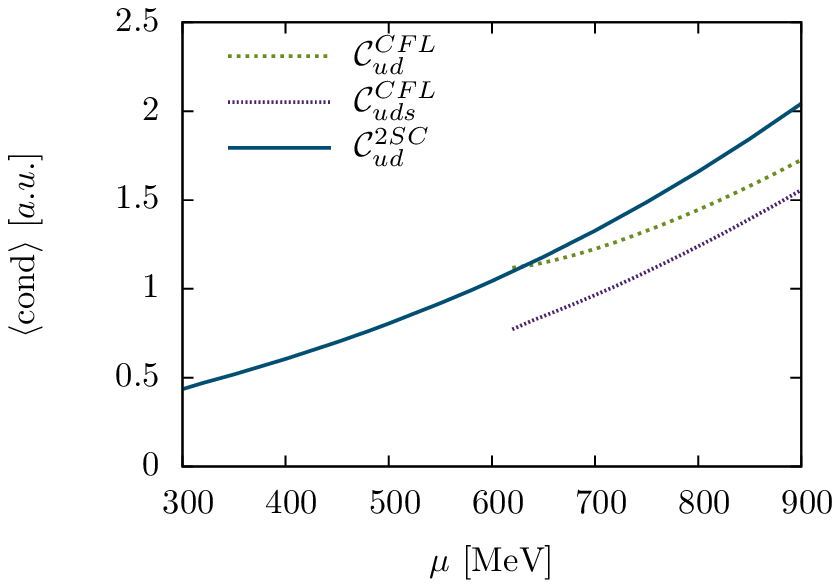}
	\caption{Dependence of 2SC and CFL condensates at $T=10$ MeV on the 
chemical potential for $m_s=30$ MeV (top) and $m_s=54$ MeV (bottom).
All condensates are given in arbitrary, but equal, units. 
}
\label{hdl_condmu}
\end{figure}

In \fig{hdl_condmu} we show the diquark condensates defined in 
\eqs{condensate} -- (\ref{eq:cond_uds}) at low temperatures, $T=10$ MeV,
as functions of the chemical potential.
As discussed in sect.~\ref{sec_struc}, our formalism allows for 
the description of both 2SC and CFL-like pairing. 
Since the condensate $\mathcal{C}_{ud}$, related to the mutual pairing
of up and down quarks, is present in both phases, we indicate explicitly
in which phase the solutions have been obtained. 
We also do so for the condensate $\mathcal{C}_{uds}$, although it is
non-zero only in the CFL phase. 
In addition, there is always a non-super\-conducting solution, 
where all diquark condensates vanish. This is not shown in the figure.

The upper and the lower panel of \fig{hdl_condmu} correspond to $m_s = 30$~MeV 
and 54~MeV, respectively. 
The qualitative behavior is similar in both cases, and even the
quantitative differences are not large.
All condensates rise with increasing chemical potential.
At lower values of $\mu$, we only find a 2SC solution of the DSE, 
whereas above a threshold of about 500 or 600~MeV, depending on $m_s$,
there is also a CFL-like solution.

\begin{figure}
    \centering
    	    \includegraphics[]{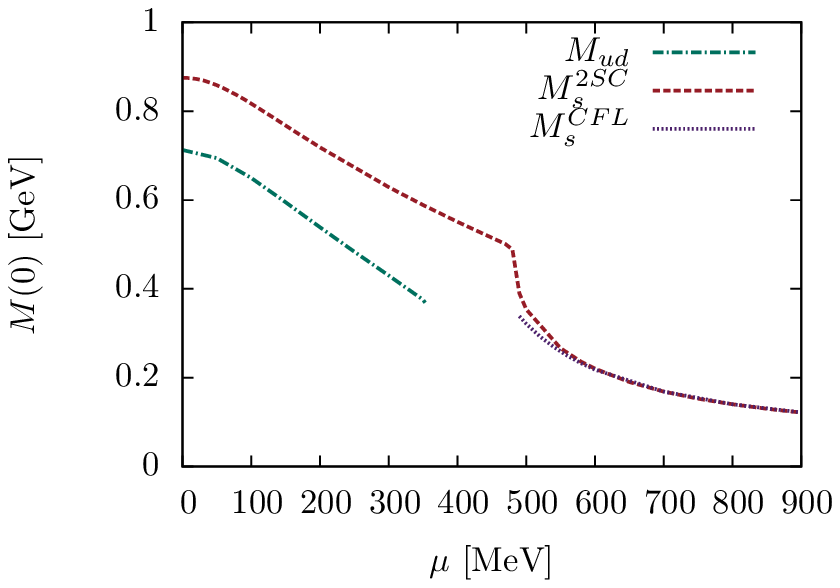}
    \centering
    	    \includegraphics[]{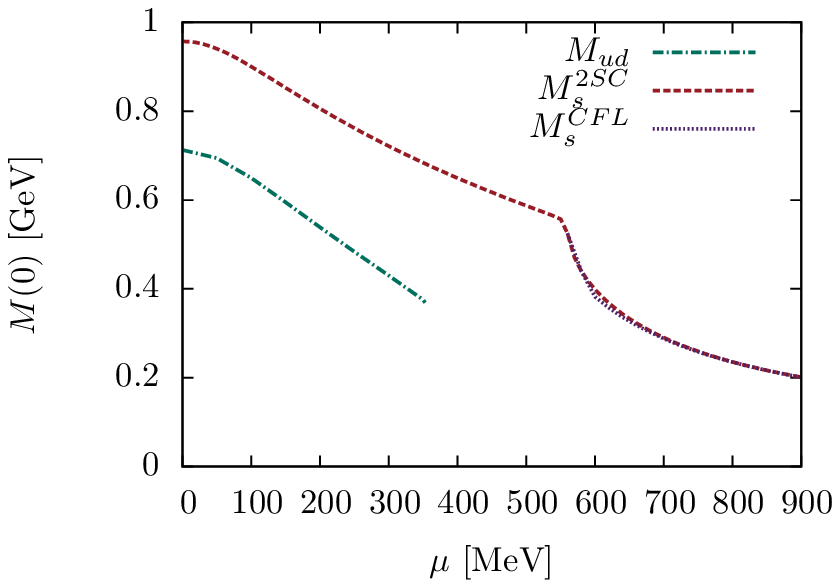}
    \caption{Dynamical quark masses
             $M(0) = B^+(0)/C^+(0)$ as functions of the
             chemical potential  at $T=10$~MeV
             for $m_s=30$ MeV (top) and $m_s=54$ MeV (bottom).
             The up- and down-quark masses are only shown for
             the non-superconducting chirally broken phase. 
             They vanish in the 2SC phase
             and are very small in the CFL phase.}
\label{hdl_mmu}
\end{figure}

This behavior is related to the $\mu$-dependence of the dynamical
quark masses, which are displayed in \fig{hdl_mmu}.
Since the normal and color-superconducting dressing functions are
mutually coupled through the DSEs, the dynamical quark masses
depend on the phase as well. 
At low $\mu$ we find a non-superconducting solution with spontaneously
broken chiral symmetry, where the up and down quarks have relatively
large masses (dash-dotted) line. 
On the other hand, there is always a 2SC solution with massless
up and down quarks. In the CFL-like solutions, the up and down quarks
are almost massless as well\footnote{In the CFL phase, the up and down
quarks have a small mass ($<$~2~MeV), which is induced by their coupling 
to the massive strange quarks. This is possible because the CFL condensates
break the chiral $SU(2)$ symmetry.
The situation is further complicated by the fact that red up, green down,
and blue strange quarks are mixed in the CFL phase (see \eq{pmcfl}).
In principle, one should therefore project on the different quasi-particle 
modes. 
However, for simplicity, we just show the $P_6$ component in \fig{hdl_mmu},
corresponding to red and green strange quarks.}.
The strange quarks, on the other hand, are quite heavy at low and 
intermediate chemical potentials. 
As a consequence, the Fermi momenta of strange and non-strange quarks
are very different, so that their mutual pairing is inhibited.
However, with increasing chemical potential, the strange quarks 
become lighter and eventually undergo a crossover to even smaller 
masses. This effect, which is seen in the 2SC phase (dashed line), 
also triggers the onset of CFL-like solutions. 
In the latter the strange-quark mass (dotted line) is slightly lower 
than in the former, but the difference is small and almost vanishes at very
high $\mu$, where the mass is dominated by perturbative effects.

As discussed earlier, the quark masses are neglected in the 
gluon polarization function in the HTL-HDL truncation.
Therefore, if there is no pairing between strange and non-strange quarks, 
the strange sector decouples from the non-strange sector.
As a consequence, the strange-quark mass is not affected by a possible
chiral phase transition of the up and down quarks.
Likewise, $\mathcal{C}_{ud}^{2SC}$ does not depend on $m_s$ and
is not influenced by the rapid change of $M_s$ in the crossover region.

\begin{figure}
	\centering
		\includegraphics[]{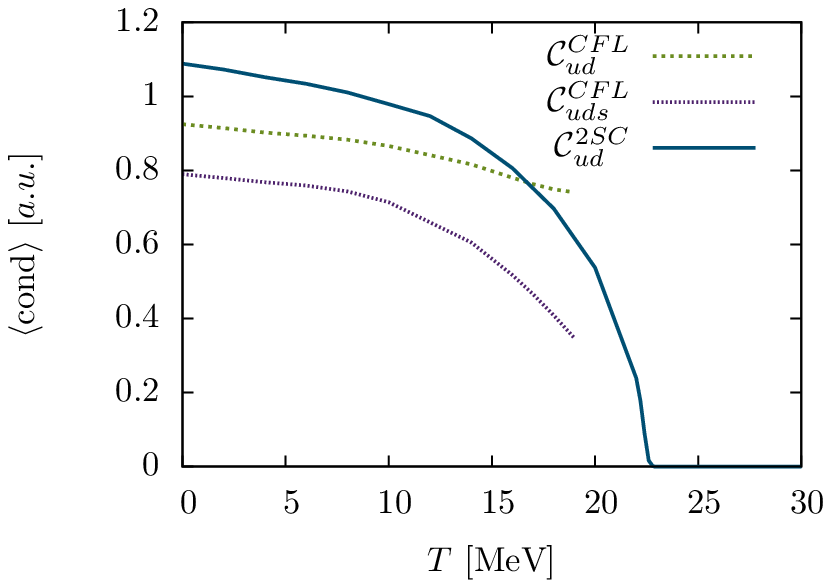}
	\centering
		\includegraphics[]{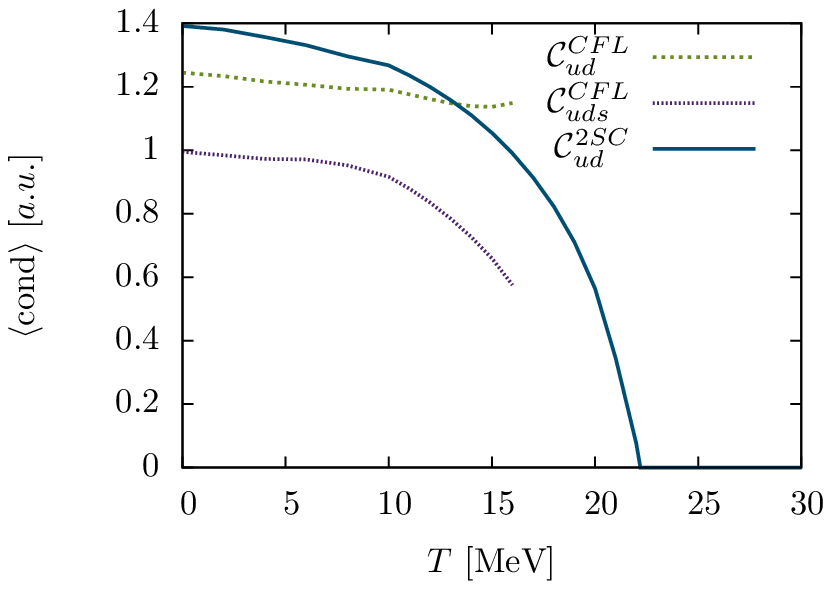}
	\caption{Temperature dependence of 2SC and CFL condensates at $\mu=580$ MeV for $m_s=30$ MeV (top) and at $\mu=680$ MeV for $m_s=54$ MeV (bottom).
The units are the same as in Fig.~\ref{hdl_condmu}.
}
\label{hdl_condT}
\end{figure}

In \fig{hdl_condT} the temperature dependence of the diquark condensates 
is shown for $m_s=30$~MeV at a chemical potential of $\mu=580$~MeV
and for $m_s=54$~MeV at $\mu=680$~MeV (lower panel).
The 2SC condensate smoothly decreases with increasing temperature
and eventually vanishes at a temperature between $20$ and $25$~MeV. 
In contrast, the CFL-like solution ceases to exist already above a lower 
critical temperature, where the condensates are still finite.
This behavior suggests that, with increasing $T$, 
the system first undergoes a first-order phase transition from the 
CFL phase to the 2SC phase, followed by a second-order phase transition 
to the normal-conducting phase.

However, if there are several solutions of the DSE at given temperature 
and chemical potential, the favored phase can only be determined by
comparing the corresponding pressures.
The pressure is equal to the effective action, which in general is given 
by 
\beq
\Gamma = \sumint\limits_p \frac{1}{2}\Tr \!\ln \SNG^{-1}(p) 
- \sumint\limits_p
\frac{1}{2}\Tr \!\lb 1\!-\! Z_2\SNG^{-1}_0(p) \SNG(p) \rb + \Gamma_2.
\eeq
Here $\Gamma_2$ denotes the two-particle irreducible part.
Our present truncation with HTL-HDL approximation corresponds to the
expression
\beq
\Gamma_2 = g^2\sumint\limits_p\sumint\limits_q\frac{1}{4} 
           \Tr\!\! \lb \Gamma^a_{\mu,0} 
           \SNG(p) D^{ab}_{\mu\nu}(p-q)\Gamma^b_\nu(p,q) \SNG(q)\rb.
\eeq
The resulting pressure is quartically divergent. 
Although finite expressions can be obtained by calculating 
pressure differences between competing phases~\cite{Nickel:2006kc},
unfortunately the results are numerically very unstable.
This is due to the fact that the convergence behavior is determined by the
high-energy tails of the integrands, which contain only little information
about the phase under consideration.

We therefore proceed as follows.
As argued in appendix \ref{app:iteration}, 
there are reasons to believe that the solutions found by solving the DSEs 
iteratively, correspond to global or local maxima of the pressure,
\textit{i.e.}, to stable or metastable solutions, but not to unstable ones. 
According to this hypothesis (which is often assumed in DSE calculations), 
this means that we always find the correct
solution in the case of second-order phase transitions.
 
For first-order phase transitions, the situation is more complicated.
In this case, there is a certain regime, where a stable and a metastable
solution of the DSE exist, which both can be found by iteration. 
The exact position of the phase transition, which manifests itself in a jump 
between the two solutions, can then only be determined by studying the 
pressure of the system.
However, because of the numerical uncertainties mentioned above, 
in practice this does not further narrow down the phase-transition region.
We therefore restrict ourselves to calculating the spinodal lines of 
the first-order region, \textit{i.e.}, the lines where the metastable 
solutions disappear. 
As these regions mostly have only a small extent, they still give a 
good estimate for the phase boundary. 

\begin{figure}
	\centering
		\includegraphics[]{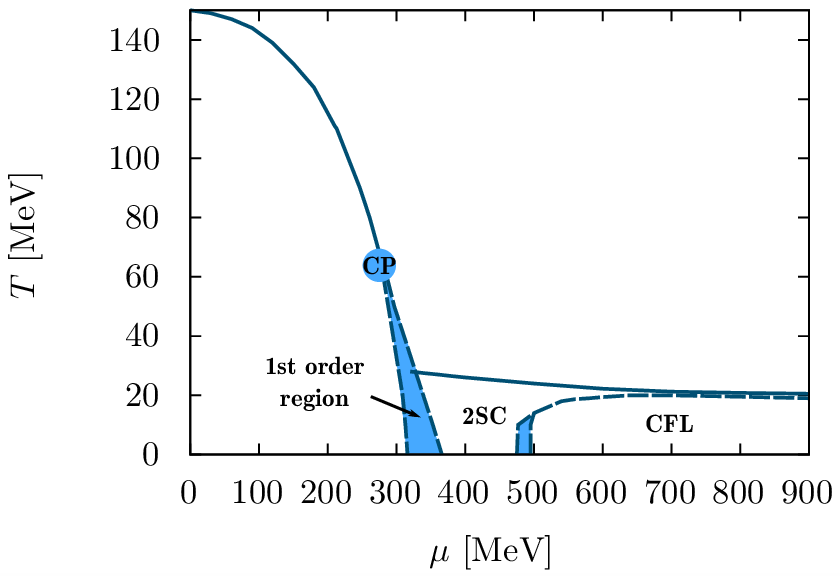}
	\centering
	    \includegraphics[]{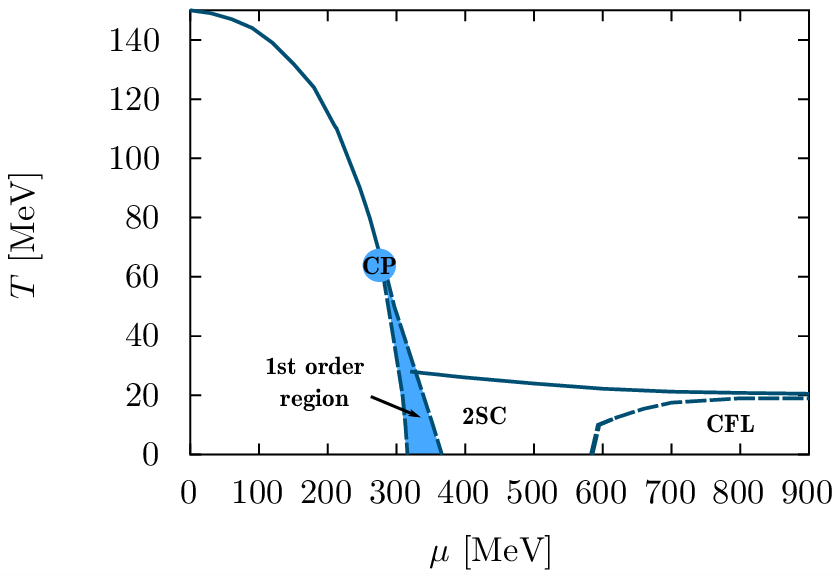}
	\caption{Phase diagram for $m_s=30$ MeV (top) and $m_s=54$ MeV 
                 (bottom). The shaded areas indicate the region of the
                  first-order phase transition, bounded by
                  spinodal lines (dashed). Solid lines indicate second-order phase transitions, CP denotes the tricritical point.}
\label{hdl_pd}
\end{figure}

The resulting phase diagrams are shown in \fig{hdl_pd}. 
Again, the upper and lower panels correspond to $m_s=30$ MeV and $54$~MeV, 
respectively. As pointed out earlier, in the HTL-HDL approximation, the
strange and non-strange sectors decouple, except for the CFL phase,
where they are coupled by the condensate. 
Therefore the only effect of $m_s$ on the phase diagram
is a shift of the CFL phase boundary.

The shaded areas indicate first-order regions bounded by spinodal lines.
The most prominent ones correspond to the chiral phase transition
from the chirally broken phase to the 2SC phase or, at somewhat
higher temperatures, to the non-superconducting chirally restored phase
(with respect to the up and down quarks).
The spinodal region and, hence, the first-order phase transition
inside this region end at a tricritical point at $T=60$~MeV and $\mu=300$~MeV.
Above this point, the chiral phase transition is of second order.

The temperature of the tricritical point is lower than in the DSE 
analysis of Ref.~\cite{Fischer:2011mz}, where it was found at 
$T=95$~MeV and $\mu=280$~MeV. The authors use a similar truncation
scheme but consider only two flavors. This also leads to a higher 
critical temperature of 180~MeV at $\mu=0$.
As shown in Ref.~\cite{Fischer:2012vc}, the inclusion of strange quarks
reduces the temperature of the critical point and shifts it to 
a slightly higher chemical potential. This is consistent with our result. 
However, since in Ref.~\cite{Fischer:2012vc} a truncation without 
the HTL-HDL approximation was used, a direct comparison of the
$2+1$-flavor results with ours is not possible.

The low-$\mu$ part of the phase diagram and the region around the
critical point are mainly shown for completeness, 
while our focus is on the color-superconducting phases at 
low temperatures and higher chemical potentials. 
Here we find a 2SC phase, followed by a CFL phase.
The latter is favored above
$\sim 500$ MeV for $m_s=30$ MeV and $\sim 600$ MeV for $m_s=54$~MeV.
The 2SC phase also extends to somewhat higher temperature than the
CFL phase (cf.\@ Fig.~\ref{hdl_condT}) and, thus, separates the latter
from the normal-conducting chirally restored phase. 
The phase transition between 2SC and restored phase is second order
and takes place around $20-30$~MeV. 
Remarkably, the critical temperature is slowly decreasing with increasing
chemical potential, despite the fact that the zero-temperature condensate 
increases (cf.~Fig.~\ref{hdl_condmu}).

As is clear from the discussion of Fig.~\ref{hdl_condmu},
the phase transition between 2SC and CFL phase at low temperature
is first order.
However, the spinodal region turns out to be very narrow and only visible for low temperatures at $m_s=30$~MeV \footnote{This means that the 2SC
solution is unstable almost everywhere where a CFL solution exists.
The 2SC condensates shown in Figs.~\ref{hdl_condmu} and \ref{hdl_condT}
have been obtained by forcing the strange condensates to be exactly
zero during the iteration. Otherwise, starting with a small non-vanishing 
strange condensate, the iteration only converges to the 2SC solution 
if we are outside the CFL regime or in the very small spinodal region.}.
Moreover, we have numerical indications that at large $\mu$ 
the phase transition is second order, \textit{i.e.}, 
the spinodal region ends again in a critical point. 
Such a behavior was also found in the NJL-model analysis of
Ref.~\cite{Buballa:2001gj}.
For $m_s=30$~MeV, the critical point seems to be somewhere around
$\mu = 900$~MeV, but it is difficult to localize its position exactly.

\section{Alternative parametrization}
\label{sec_vacparam} 

As pointed out before, the parametrization of the quark-gluon vertex function 
used so far, yields too large values for the pion decay constant and the chiral condensate 
in vacuum. 
In this section, we will discuss this in more detail. 
It raises the question to what extent the results presented in the previous section, 
and in particular the existence of a 2SC phase at intermediate chemical potentials,
are consequences of a potentially too strong attraction in the infrared. 
We will therefore determine an alternative parametrization by fitting
vacuum observables instead of the chiral critical temperature.
Afterwards, we present the resulting phase diagram.

\subsection{Chiral condensate and pion decay constant}
\label{sec:cc_pd}

For massless quarks the renormalization dependent chiral condensate can be calculated with \eq{eq:qq_cond}. For the parametrization used so far this yields the vacuum value 
$\langle \bar u u \rangle(\nu) =-0.14$~GeV$^3$ at our renormalization point $\nu = 100$~GeV
(cf.~\fig{fig:condT}).
However, for comparison with literature values, we need the renormalization-point independent
condensate. Employing the operator product expansion, this quantity can be extracted
by fitting the asymptotic behavior of the quark-mass function~\cite{Williams:2007ey}
\beq
M(p) \overset{p\rightarrow \infty}{=}-\langle \bar u u \rangle\frac{2\pi^2\gamma_m}{3p^2}\lb\frac{1}{2}\log(p^2/\bar\Lambda^2)\rb^{\gamma_m-1}
\eeq
with the anomalous dimension $\gamma_m=12/(11N_c-2N_f)$.
In a full calculation, the parameter $\bar\Lambda$ would correspond to $\Lambda_{QCD}$. 
As we solve a truncated system, its value can be different and we it as a fit parameter as well. 
We then get a value of $\langle\bar u u \rangle = -(425\mbox{~MeV})^3$ for the 
renormalization-point independent condensate. 

The pion decay constant can be estimated with the Pagels-Stokar formula \cite{Pagels:1979hd,Roberts:1994dr}
\bseq
f_\pi^2 = &\frac{N_c}{4\pi^2}\int_0^\infty dp^2p^2 \frac{Z_2 A^{-1}(p^2) M(p^2)}{\lb p^2+M^2(p^2)\rb^2}\\ &\left(M(p^2) - \frac{p^2}{2}\frac{dM(p^2)}{dp^2}\right)~.
\eseq
For the parametrization used so far, we get $f_\pi = 127$~MeV.

\subsection{Vacuum fit}

Hence, both, the chiral condensate and the pion decay constant, are too large compared with 
realistic values $\langle\bar u u \rangle \approx -(250\mbox{~MeV})^3$ and
$f_\pi=92.4$~MeV (or $f_\pi = 88$~MeV in the chiral limit).
This deviation suggests that the infrared enhancement of the quark-gluon vertex, which is
mainly determined by the parameters $d_1$ and $d_2$ in \eq{vertex1}, is possibly too strong.
Indeed, we can easily obtain a better fit of $f_\pi$ be choosing a smaller value of $d_1$.
It turns out, however, that in order to get sufficiently strong reduction of the chiral condensate,
we also have to modify the scale of the logarithmic running and the parametrization of the gluon 
dressing functions.

In fact, the lattice calculation of the gluon propagator only provides data up to a few GeV,
leaving some freedom for the parameters in  \eq{eq:ZYM}.
In the new parametrization, we take $a_T = a_L=1.01$, $b_T=b_L=0.8$, $c=16.3$~GeV$^2$,
$\alpha(\nu)=0.727$ and $\Lambda=1.160$~GeV. 
For the logarithmic running in the expression $\ln(k^2/\bar\Lambda^2+1)$ we choose a smaller 
scale of $\bar\Lambda=0.6$~GeV. 
There is no need to introduce temperature dependent parameters, since we will stay 
below $T\approx 100$ MeV, where the lattice data are almost temperature independent. 
 
For the vertex dressing we use the parametrization
\bseq
&\Gamma(k) = Z_2\tilde Z_3\frac{\Lambda^2}{k^2+\Lambda^2}\\
&\lb\lb\frac{d_1}{d_2+k^2}\rb^2 + \frac{k^2}{\Lambda^2}\left(\frac{\beta_0\alpha(\nu)\ln(k^2/\bar\Lambda^2+1)}{4\pi}\right)^{2\delta} \rb
\eseq
with $d_1=3.15$~GeV$^2$, $d_2=0.5$~GeV$^2$, and $\Lambda$ and $\bar\Lambda$ as
specified above. 

With this choice we obtain the realistic vacuum values
$f_\pi=87.5$ MeV and $\langle\bar u u \rangle= -(260\mbox{~MeV})^3$. 
The latter corresponds to a renormalization-point dependent condensate of 
$\=-0.035$~GeV$^3$ at $\nu = 100$~MeV. 
We can then employ the Gell-Mann--Oakes--Renner relation,
$f_\pi^2 m_\pi^2 = -2m_{u,d}\langle \bar u u \rangle$ to estimate the value of
the up and down quarks at this scale. Taking $m_\pi = 135$~MeV, this yields
$m_{u,d} = 2$~MeV. Finally, since the mass ratio of strange and non-strange 
quark is $m_s/m_{u,d} = 27.5$ \cite{Beringer:1900zz}, we conclude that the
strange-quark mass is about $55$~MeV.
In the following, we will therefore restrict ourselves to our larger value of the strange-quark
mass, $m_s(\nu = 100\,\mathrm{GeV}) = 54$~MeV, and take chiral up and down quarks, 
as before.

\subsection{Phase diagram}
\label{sec:cc_pd_res}

\begin{figure}
	\centering
		\includegraphics[]{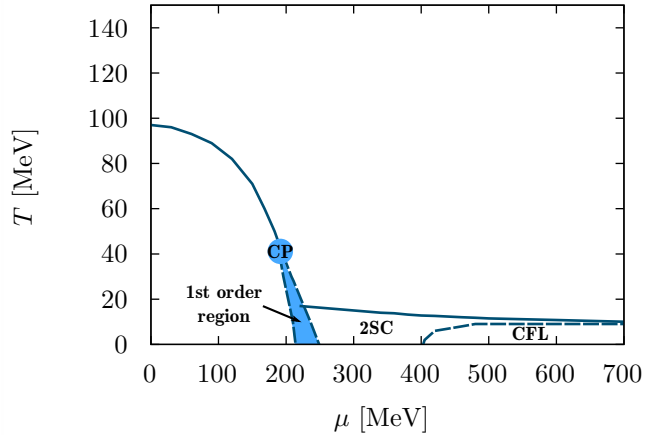}
	\caption{Phase diagram for a parametrization fixed by vacuum quantities.}
\label{fig:pdvac}
\end{figure}

The phase diagram obtained with this parametrization is shown in \fig{fig:pdvac}.
To first approximation, it looks similar to the phase diagrams in \fig{hdl_pd}
but with all critical temperatures and chemical potentials reduced by about 30\%.
In particular the chiral phase transition temperature at $\mu=0$ is way below the 
lattice value of about 150~MeV. Also the critical chemical potential at $T=0$ is 
totally unrealistic, as it below the nuclear-matter point, $\mu_{nm} = 308$~MeV. 
We will come back to this problem in the Conclusions below. 

Our main result at this point is that the qualitative phase structure is very similar 
as for our previous parame\-trization. In particular, we still find a region with a stable 
2SC phase.

\section{Conclusions and outlook}
\label{sec_conclusions}

We have investigated color-superconducting phases in QCD with Dyson-Schwinger 
equations in Landau gauge.
For the gluon propagator we used a recent parametrization of quenched 
lattice data and included quark-loop effects perturbatively in
a hard-thermal-loop--hard-dense-loop approximation.
For the vertex we used a phenomenological ansatz constrained by the known
asymptotic behavior in the ultraviolet, while the infrared strength is
fitted to obtain a chiral phase-transition temperature of 150~MeV at
$\mu = 0$.

We have studied 2SC and CFL-like phases for $N_f=2+1$ flavors with chiral 
up and down quarks, and massive strange quarks. We find a dominant CFL phase 
at high chemical potentials and low temperatures
while there is always a thin 2SC band at higher temperature, 
separating the CFL phase from the non-superconducting chirally restored phase.
The transition to the latter takes place at critical temperatures
around $T=20-30$ MeV.

Our most important result is that we also find a stable 2SC phase at small 
temperatures and intermediate chemical potential, $\mu\lesssim 500$ MeV.
This is in strong contrast to the results of Ref.~\cite{Nickel:2006kc}
where the CFL phase was found to be favored practically all the way 
down to the hadronic phase at zero temperature. 

The difference should therefore mainly be attributed to the improved
gluon propagators and vertices, which we have used.
This interaction is effectively stronger than those used in 
Ref.~\cite{Nickel:2006kc}. As a consequence, the strange quarks stay 
heavy up to higher chemical potentials, inhibiting their pairing with 
the non-strange quarks.

On the other hand, the vacuum values for the chiral condensate and the pion decay constant
obtained with this parametrization are much too large.  
We have therefore performed calculations with an alternative parametrization, where these
vacuum quantities have been fitted to their empirical values.  
We found that the resulting phase structure stays qualitatively unchanged and 
the stable 2SC region persists. 
The critical temperatures and chemical potentials are, however, too small.

This observation that we were not able to obtain a good vacuum fit and a reasonable critical
temperature simultaneously is clearly the most unsatisfactory feature of the employed
truncation scheme. In fact, the HTL-HDL approximation neglects quark-mass and pairing-gap 
effects on the quark-polarization loop in the gluon propagator and therefore overestimates
the gluon screening in the chirally broken and color superconducting phases. 
Therefore we are currently working on an improved scheme, 
where the dressed Nambu-Gorkov propagators from the quark DSE
are self-consistently used to evaluate the polarization loop.
Preliminary results indicate that this leads to quantitative changes, 
bridging the gap between the parametrizations fitted to vacuum quantities or
to the chiral phase transition, respectively. 
In particular, when the parameters are fitted to the chiral critical temperature $T=150$~MeV,
the dressed quark masses and the pion decay constant are considerably lower than in
HTL-HDL approximation.
Also, the critical temperatures for CSC phases increase. However, the structure
of the phase diagram is not altered qualitatively and the stable 2SC phase persists \cite{Mueller:2013ip}. 

Further possible improvements concern the quark-gluon vertex, which is
only constrained in the UV so far, while for the IR behavior we used 
a phenomenological model ansatz. 
It has been studied on the lattice \cite{Skullerud:2001bu} and also with Dyson-Schwinger equations in semi-perturbative truncations \cite{Alkofer:2008tt,Fischer:2009jm}. Recently the coupled system of quark DSE and vertex DSE has been investigated self-consistently in a truncated version in vacuum \cite{Hopfer:2013np} and it is desirable to include non-abelian vertex structures and finally to implement a truncated vertex DSE self-consistently. However, studies of the vertex DSE in medium are very difficult and have not been performed so far. 

In the present paper, we have considered a single, flavor independent
chemical potential. These studies should be extended to electrically
neutral matter in beta equilibrium, as relevant for compact stars. 
Such an analysis has been performed in Ref.~\cite{Nickel:2008ef} for
zero temperature using the ``old'' gluonic interaction. 
Since electric neutrality splits the Fermi surfaces of up and down
quarks, it disfavors the 2SC phase~\cite{Alford:2002kj}.
Therefore, it would be interesting to see, whether this phase survives
when the improved interaction is used.

Another interesting question is, whether the QCD phase diagram contains
inhomogeneous phases.
Phases with spatially varying chiral condensates have been found
in effective models~\cite{Nickel:2009wj,Broniowski:2011ef}
and in large-$N_c$ QCD with weak-coupling 
methods~\cite{Deryagin:1992rw,Shuster:1999tn}. 
It would be exciting to study these phases with DSEs as well.
This also includes inhomogeneous color 
superconductors~\cite{Bowers:2002xr,Casalbuoni:2003wh,Anglani:2013gfu},
which may exist when electric neutrality is enforced. 

Finally, it is an important, but difficult, task to calculate the pressure 
in the various phases, which would allow for a more precise determination 
of the first-order phase boundaries. 
Although an analytical expression for the pressure is given by the
effective potential, its numerical evaluation was so far hindered by the
poor convergence of the integrals. 
A solution of this problem would constitute a major breakthrough.

\section{Acknowledgements} 
We would like to thank Christian Fischer and Jan L\"ucker for helpful discussions and comments. D.M. was supported by BMBF under contract 06DA9047I and by the Helmholtz Graduate School for Hadron and Ion Research.
M.B. and J.W. acknowledge partial support by the Helmholtz International
Center for FAIR and by the Helmholtz Institute EMMI.

\appendix

\section{Color-flavor structure}\label{cfmatrices}

The matrices $P_i$ and $M_i$, which parametrize the color-flavor structure of 
the quark propagator and self-energies, Eqs.~(\ref{eq:SPTM}) and 
(\ref{eq:SigmaPPhiM}), are constructed such that they yield a closed set of 
self-consistency equations when inserted into \eq{dseng}.
Additional restrictions come from the requirement that
the resulting propagator must be consistent with the symmetries 
of the considered phase.

In the most general case considered here, 
the CFL-like pairing pattern with finite strange-quark mass, 
the residual color-flavor symmetry is   
$SU_{c+V}(2) \otimes U_{c+V}(1)$, generated by $\tau_a - \lambda_a^T$
with $a = 1,2,3$, and 8.
One then finds that the sets $\{P_i\}$ and $\{M_i\}$ consist of seven 
matrices each~\cite{Nickel:2006kc}.
In the color-flavor basis\\
$\lbrace(r,u),(g,d),(b,s),(r,d),(g,u),(r,s),(b,u),(g,s),(b,d)\rbrace$\\
these are given by
\beq\label{pmcfl}
P_i = \left(\begin{smallmatrix}
\delta_{i,1}+\delta_{i,2} & \delta_{i,2} & \delta_{i,4}&&&&&&\\
\delta_{i,2} & \delta_{i,1}+\delta_{i,2} & \delta_{i,4}&&&&&&\\
\delta_{i,5} & \delta_{i,5} & \delta_{i,3}&&&&&&\\
&&&\delta_{i,1}&&&&\\
&&&&\delta_{i,1}&&&&\\
&&&&&\delta_{i,6}&&&\\
&&&&&&\delta_{i,7}&&\\
&&&&&&&\delta_{i,6}&\\
&&&&&&&&\delta_{i,7}\\
\end{smallmatrix}\right),
\eeq

\beq\label{mmcfl}
M_i = \left(\begin{smallmatrix}
\delta_{i,1}+\delta_{i,2} & \delta_{i,2} & \delta_{i,4}&&&&&\\
\delta_{i,2} & \delta_{i,1}+\delta_{i,2} & \delta_{i,4}&&&&&\\
\delta_{i,5} & \delta_{i,5} & \delta_{i,3}&&&&&\\
&&&&\delta_{i,1}&&&\\
&&&\delta_{i,1}&&&&\\
&&&&&\delta_{i,7}&&\\
&&&&\delta_{i,6}&&&\\
&&&&&&&\delta_{i,7}\\
&&&&&&\delta_{i,6}&\\
\end{smallmatrix}\right),
\eeq
where we have slightly modified the notation of 
Ref.~\cite{Nickel:2006kc}.\footnote{
Our  $P_6$ and $M_6$ correspond to  $P_7$ and $M_7$ 
in Ref.~\cite{Nickel:2006kc} and 
our $P_7$ and $M_7$ correspond to $P_8$ and $M_8$,  
while there are no projectors $P_6$ and $M_6$ in that reference.
}

More symmetric phases are contained in this parametri\-zation as special
cases:

In the non-superconducting phase all anomalous self-energy components vanish,
$\phi_i^+ = 0$,
and the normal self-energy is diagonal in color and flavor, 
\textit{i.e.}, $\Sigma_i^+ = 0$ for $i=2,4,5$. 
The remaining self-energy components are equal for up and down quarks,
but can be different for strange quarks,
\textit{i.e.}, $\Sigma_1^+ = \Sigma_7^+$ and $\Sigma_3^+ = \Sigma_6^+$.

In the 2SC phase, only red and green up and down quarks are paired,
described by $\phi_1^+ = - \phi_2^+$, whereas all other components of the
 anomalous self-energy vanish.
Accordingly, the normal self-energy depends on whether it corresponds to
paired (red or green) up or down quarks ($\Sigma_1^+$), 
unpaired (blue) up or down quarks ($\Sigma_7^+$),
or to strange quarks ($\Sigma_3^+=\Sigma_6^+$), 
while the non-diagonal components vanish again. 

In the CFL phase with three massless flavors, the residual symmetry
is $SU_{c+V}(3)$, generated by $\tau_a - \lambda_a^T$
with $a = 1,\dots,8$, and the quasi-particle spectrum consists of an
octet and a singlet~\cite{Alford:1998mk}.
Therefore there are only two independent color-flavor components of 
the anomalous self-energy~\cite{Alford:1999pa},
$\phi_1^+ = \phi_6^+ = \phi_7^+$, $\phi_2^+ = \phi_4^+ = \phi_5^+$, and 
 $\phi_3^+ = \phi_1^+ + \phi_2^+$.
Analogous relations hold for the normal self-energies.

\section{A remark on the iterative procedure}
\label{app:iteration}

Throughout this work we use a fixed point iteration to obtain the 
self-consistent solutions of the DSE. 
Schematically, this can be formulated as an iteration
\beq\label{cond_it}
x_{i+1} = \varphi(x_i)
\eeq
with the fixed points $x^*=\varphi(x^*)$, which correspond to the 
solutions of the DSE.

In general, however, not every solution can be found in this way, 
but the iteration only converges if it is contracting, \textit{i.e.}, if 
\beq
\label{iterstab}
\frac{\vert x_{i+1}-x^* \vert}{\vert x_{i}-x^* \vert}=\frac{\vert\varphi(x_i)-\varphi(x^*) \vert}{\vert x_{i}-x^* \vert} \overset{!}{\leq} L
\eeq
with a constant $L<1$. 
This expression is just the discretized derivative $\vert \varphi'(x^*) \vert$.

The phase diagrams and in particular the spinodal lines presented in 
Sec.~\ref{sec_results} are based on the assumption that the numerically
stable solutions of the DSE, \textit{i.e.}, those solutions for which the iteration
converges, also correspond to thermodynamically stable or metastable 
solutions, \textit{i.e.}, to global or local minima of the thermodynamic potential.

This conjecture is motivated by the observation that it holds in the
NJL model, where it can be shown analytically.
To this end, we consider the standard NJL model, defined by the
Lagrangian~\cite{Nambu:1961fr}
\beq
    {\cal L}_\mathit{NJL} 
    = \bar\psi(i\slashed{\partial} -m)\psi 
     + G\left[ (\bar\psi\psi)^2 + (\bar\psi i\gamma_5\vec\tau\psi)^2\right]
\eeq
with bare quark mass $m$ and a positive four-point coupling constant $G$. 
The NJL mean-field thermodynamic potential in vacuum is given by 
(see, \textit{e.g.}, Ref.~\cite{Buballa:2003qv})
\beq
\Omega
= \frac{(M-m)^2}{4G} - 2N_cN_f\int\frac{d^3q}{(2\pi)^3} \sqrt{\vec q\,^2+M^2},
\eeq
which depends on the dressed quark mass $M$. Additionally a regularization 
needs to be specified, as the integral is divergent.

The actual value of $M$ is determined by minimizing $\Omega$. 
This leads to the stationarity condition
\beq
\label{NJL_gap}
\frac{\delta\Omega}{\delta M} = \frac{M-\varphi(M)}{2G} \overset{!}{=} 0
\eeq
with
\beq
\label{phi_M}
    \varphi(M) = 
     m + 4G N_cN_f \int\frac{d^3q}{(2\pi)^3} \frac{M}{\sqrt{\vec q^2+M^2}}\,.
\eeq
Obviously, \eq{NJL_gap}
is equivalent to the DSE (``gap equation'') 
$M = \varphi(M)$, which has the solutions $M^*$.
The physical stability of these solutions can then be checked by 
investigating the second derivative
\beq
\label{pot_minmax}
\left.\frac{\delta^2\Omega}{\delta M^2}\right\vert_{M=M^*} =
\frac{1}{2G} \big(1- \varphi'(M^*)\big)
\eeq
with $\varphi'(M) = \frac{d\varphi(M)}{dM}$.
This means, the solution corresponds to a minimum if $\varphi'(M^*)<1$
and to a maximum if  $\varphi'(M^*)>1$.

Taking the derivative of \eq{phi_M}, one obtains
\beq
\varphi'(M)
=
4GN_cN_f\int\frac{d^3q}{(2\pi)^3} 
\frac{\vec q\,^2}{(\vec q\,^2+ M^{2})^{3/2}}
\eeq
which is positive for all values of $M$. 
Therefore, we have
\beq
|\varphi'(M^*)| \;
\begin{cases}
< 1 \quad \text{for a minimum of} \;\; \Omega
\\
> 1 \quad \text{for a maximum of} \;\; \Omega
\end{cases}
\eeq
Comparing this with \eq{iterstab}, we see immediately 
that the maxima of the potential correspond to numerical unstable iterative 
solutions, while minima are numerically stable. This allows in principle to 
find all minima of the potential by iterating the gap equation. 
It can also be checked easily that the introduction of finite temperature 
and chemical potential does not change these results. 

For QCD DSEs, the situation is much more complicated as we have functional 
derivatives and an in principle infinite dimensional system. 
Therefore we are not able to provide an analogous analytic argument for 
the relation between physical and numerical stability.
However, we expect this relation still to be valid. 


\bibliography{literature}

\begin{thebibliography}{10}

\bibitem{BraunMunzinger:2009zz}
P.~{Braun-Munzinger} and J.~{Wambach},
\newblock Reviews of Modern Physics {\bf 81}, 1031 (2009).

\bibitem{Fukushima:2010bq}
K.~{Fukushima} and T.~{Hatsuda},
\newblock Reports on Progress in Physics {\bf 74}, 014001 (2011),
  \href{http://arxiv.org/abs/1005.4814}{{\ttfamily arXiv:1005.4814 [hep-ph]}}.

\bibitem{Bazavov:2011nk}
A.~{Bazavov} {\em et~al.},
\newblock \prd {\bf 85}, 054503 (2012),
  \href{http://arxiv.org/abs/1111.1710}{{\ttfamily arXiv:1111.1710 [hep-lat]}}.

\bibitem{Borsanyi:2010bp}
S.~{Bors{\'a}nyi} {\em et~al.},
\newblock Journal of High Energy Physics {\bf 9}, 73 (2010),
  \href{http://arxiv.org/abs/1005.3508}{{\ttfamily arXiv:1005.3508 [hep-lat]}}.

\bibitem{deForcrand:2010he}
P.~{de Forcrand} and O.~{Philipsen},
\newblock Physical Review Letters {\bf 105}, 152001 (2010),
  \href{http://arxiv.org/abs/1004.3144}{{\ttfamily arXiv:1004.3144 [hep-lat]}}.

\bibitem{Endrodi:2011gv}
G.~{Endr{\H o}di}, Z.~{Fodor}, S.~D. {Katz}, {Szab{\'o}}, and {K.~K.},
\newblock Journal of High Energy Physics {\bf 4}, 1 (2011),
  \href{http://arxiv.org/abs/1102.1356}{{\ttfamily arXiv:1102.1356 [hep-lat]}}.

\bibitem{Kaczmarek:2011zz}
O.~{Kaczmarek} {\em et~al.},
\newblock \prd {\bf 83}, 014504 (2011),
  \href{http://arxiv.org/abs/1011.3130}{{\ttfamily arXiv:1011.3130 [hep-lat]}}.

\bibitem{Karsch:2011yq}
F.~{Karsch}, B.-J. {Schaefer}, M.~{Wagner}, and J.~{Wambach},
\newblock Physics Letters B {\bf 698}, 256 (2011),
  \href{http://arxiv.org/abs/1009.5211}{{\ttfamily arXiv:1009.5211 [hep-ph]}}.

\bibitem{Borsanyi:2012cr}
S.~{Bors{\'a}nyi} {\em et~al.},
\newblock Journal of High Energy Physics {\bf 8}, 53 (2012),
  \href{http://arxiv.org/abs/1204.6710}{{\ttfamily arXiv:1204.6710 [hep-lat]}}.

\bibitem{Rajagopal:2000wf}
K.~{Rajagopal} and F.~{Wilczek},
\newblock {The Condensed Matter Physics of QCD},
\newblock in {\em At the frontier of particle physics, vol. 3}, edited by
  M.~Shifman, pp. 2061--2151, World Scientific, Singapore, 2001,
  \href{http://arxiv.org/abs/arXiv:hep-ph/0011333}{{\ttfamily
  arXiv:hep-ph/0011333}}.

\bibitem{Alford:2001dt}
M.~{Alford},
\newblock Annual Review of Nuclear and Particle Science {\bf 51}, 131 (2001),
  \href{http://arxiv.org/abs/arXiv:hep-ph/0102047}{{\ttfamily
  arXiv:hep-ph/0102047}}.

\bibitem{Schafer:2003vz}
T.~{Sch{\"a}fer},
\newblock Proc. of the BARC workshop on Quarks and Mesons, Mumbai  (2003),
  \href{http://arxiv.org/abs/arXiv:hep-ph/0304281}{{\ttfamily
  arXiv:hep-ph/0304281}}.

\bibitem{Rischke:2003mt}
D.~H. {Rischke},
\newblock Progress in Particle and Nuclear Physics {\bf 52}, 197 (2004),
  \href{http://arxiv.org/abs/arXiv:nucl-th/0305030}{{\ttfamily
  arXiv:nucl-th/0305030}}.

\bibitem{Buballa:2003qv}
M.~{Buballa},
\newblock \physrep {\bf 407}, 205 (2005),
  \href{http://arxiv.org/abs/arXiv:hep-ph/0402234}{{\ttfamily
  arXiv:hep-ph/0402234}}.

\bibitem{Shovkovy:2004me}
I.~A. {Shovkovy},
\newblock Foundations of Physics {\bf 35}, 1309 (2005),
  \href{http://arxiv.org/abs/arXiv:nucl-th/0410091}{{\ttfamily
  arXiv:nucl-th/0410091}}.

\bibitem{Alford:2007xm}
M.~G. {Alford}, A.~{Schmitt}, K.~{Rajagopal}, and T.~{Sch{\"a}fer},
\newblock Reviews of Modern Physics {\bf 80}, 1455 (2008),
  \href{http://arxiv.org/abs/0709.4635}{{\ttfamily arXiv:0709.4635 [hep-ph]}}.

\bibitem{Alford:1998mk}
M.~{Alford}, K.~{Rajagopal}, and F.~{Wilczek},
\newblock Nuclear Physics B {\bf 537}, 443 (1999),
  \href{http://arxiv.org/abs/arXiv:hep-ph/9804403}{{\ttfamily
  arXiv:hep-ph/9804403}}.

\bibitem{Schafer:1999fe}
T.~{Sch{\"a}fer},
\newblock Nuclear Physics B {\bf 575}, 269 (2000),
  \href{http://arxiv.org/abs/arXiv:hep-ph/9909574}{{\ttfamily
  arXiv:hep-ph/9909574}}.

\bibitem{Shovkovy:1999mr}
I.~A. {Shovkovy} and L.~C.~R. {Wijewardhana},
\newblock Physics Letters B {\bf 470}, 189 (1999),
  \href{http://arxiv.org/abs/arXiv:hep-ph/9910225}{{\ttfamily
  arXiv:hep-ph/9910225}}.

\bibitem{Buballa:2001gj}
M.~{Buballa} and M.~{Oertel},
\newblock Nuclear Physics A {\bf 703}, 770 (2002),
  \href{http://arxiv.org/abs/arXiv:hep-ph/0109095}{{\ttfamily
  arXiv:hep-ph/0109095}}.

\bibitem{Abuki:2005ms}
H.~{Abuki} and T.~{Kunihiro},
\newblock Nuclear Physics A {\bf 768}, 118 (2006),
  \href{http://arxiv.org/abs/arXiv:hep-ph/0509172}{{\ttfamily
  arXiv:hep-ph/0509172}}.

\bibitem{Ruester:2006aj}
S.~B. {R{\"u}ster}, V.~{Werth}, M.~{Buballa}, I.~A. {Shovkovy}, and D.~H.
  {Rischke},
\newblock {Phase Diagram of Neutral Quark Matter at Moderate Densities},
\newblock in {\em Pairing in Fermionic Systems: Basic Concepts and Modern
  Applications}, edited by A.~{Sedrakian}, J.~W. {Clark}, and M.~{Alford},
  p.~63, World Scientific Publishing Co, 2006.

\bibitem{Blaschke:2005uj}
D.~{Blaschke}, S.~{Fredriksson}, H.~{Grigorian}, A.~M. {{\"O}zta{\c s}}, and
  F.~{Sandin},
\newblock \prd {\bf 72}, 065020 (2005),
  \href{http://arxiv.org/abs/arXiv:hep-ph/0503194}{{\ttfamily
  arXiv:hep-ph/0503194}}.

\bibitem{Fischer:2009gk}
C.~S. {Fischer} and J.~A. {M{\"u}ller},
\newblock \prd {\bf 80}, 074029 (2009),
  \href{http://arxiv.org/abs/0908.0007}{{\ttfamily arXiv:0908.0007 [hep-ph]}}.

\bibitem{Fischer:2010fx}
C.~S. {Fischer}, A.~{Maas}, and J.~A. {M{\"u}ller},
\newblock European Physical Journal C {\bf 68}, 165 (2010),
  \href{http://arxiv.org/abs/1003.1960}{{\ttfamily arXiv:1003.1960 [hep-ph]}}.

\bibitem{Fischer:2011mz}
C.~S. {Fischer}, J.~{L{\"u}cker}, and J.~A. {M{\"u}ller},
\newblock Physics Letters B {\bf 702}, 438 (2011),
  \href{http://arxiv.org/abs/1104.1564}{{\ttfamily arXiv:1104.1564 [hep-ph]}}.

\bibitem{Fischer:2012vc}
C.~S. {Fischer} and J.~{L{\"u}cker},
\newblock Physics Letters B {\bf 718}, 1036 (2013),
  \href{http://arxiv.org/abs/1206.5191}{{\ttfamily arXiv:1206.5191 [hep-ph]}}.

\bibitem{Nickel:2006vf}
D.~{Nickel}, J.~{Wambach}, and R.~{Alkofer},
\newblock \prd {\bf 73}, 114028 (2006),
  \href{http://arxiv.org/abs/arXiv:hep-ph/0603163}{{\ttfamily
  arXiv:hep-ph/0603163}}.

\bibitem{Nickel:2006kc}
D.~{Nickel}, R.~{Alkofer}, and J.~{Wambach},
\newblock \prd {\bf 74}, 114015 (2006),
  \href{http://arxiv.org/abs/arXiv:hep-ph/0609198}{{\ttfamily
  arXiv:hep-ph/0609198}}.

\bibitem{Nickel:2008ef}
D.~{Nickel}, R.~{Alkofer}, and J.~{Wambach},
\newblock \prd {\bf 77}, 114010 (2008),
  \href{http://arxiv.org/abs/0802.3187}{{\ttfamily arXiv:0802.3187 [hep-ph]}}.

\bibitem{Nambu:1960tm}
Y.~{Nambu},
\newblock Physical Review {\bf 117}, 648 (1960).

\bibitem{Gorkov}
L.~P. Gorkov,
\newblock JETP {\bf 36(9)}, 1364 (1959).

\bibitem{Pisarski:1999av}
R.~D. {Pisarski} and D.~H. {Rischke},
\newblock \prd {\bf 60}, 094013 (1999),
  \href{http://arxiv.org/abs/arXiv:nucl-th/9903023}{{\ttfamily
  arXiv:nucl-th/9903023}}.

\bibitem{Ball:1980ay}
J.~S. {Ball} and T.-W. {Chiu},
\newblock \prd {\bf 22}, 2542 (1980).

\bibitem{Maas:2004se}
A.~{Maas}, J.~{Wambach}, B.~{Gr{\"u}ter}, and R.~{Alkofer},
\newblock European Physical Journal C {\bf 37}, 335 (2004),
  \href{http://arxiv.org/abs/arXiv:hep-ph/0408074}{{\ttfamily
  arXiv:hep-ph/0408074}}.

\bibitem{Maas:2005hs}
A.~{Maas}, J.~{Wambach}, and R.~{Alkofer},
\newblock European Physical Journal C {\bf 42}, 93 (2005),
  \href{http://arxiv.org/abs/arXiv:hep-ph/0504019}{{\ttfamily
  arXiv:hep-ph/0504019}}.

\bibitem{Cucchieri:2007ta}
A.~{Cucchieri}, A.~{Maas}, and T.~{Mendes},
\newblock \prd {\bf 75}, 076003 (2007),
  \href{http://arxiv.org/abs/arXiv:hep-lat/0702022}{{\ttfamily
  arXiv:hep-lat/0702022}}.

\bibitem{Bellac:2000b}
M.~{Le Bellac},
\newblock {\em {Thermal Field Theory}} (Cambridge University Press, 2000).

\bibitem{Mueller:2013ip}
D.~M{\"u}ller, M.~Buballa, and J.~Wambach,
\newblock {in preparation} .

\bibitem{Beringer:1900zz}
J.~Beringer and et~al.,
\newblock \prd {\bf 86}, 010001 (2012).

\bibitem{Durr:2010vn}
{Budapest-Marseille-Wuppertal Collaboration} {\em et~al.},
\newblock Physics Letters B {\bf 701}, 265 (2011),
  \href{http://arxiv.org/abs/1011.2403}{{\ttfamily arXiv:1011.2403 [hep-lat]}}.

\bibitem{Chetyrkin:1999pq}
K.~G. {Chetyrkin} and A.~{Retey},
\newblock {Nuclear Physics B} {\bf 583}, 3 (2000),
  \href{http://arxiv.org/abs/arXiv:hep-ph/9910332}{{\ttfamily
  arXiv:hep-ph/9910332}}.

\bibitem{Williams:2007ey}
R.~{Williams}, C.~S. {Fischer}, and M.~R. {Pennington},
\newblock arXiv:0704.2296 [hep-ph]  (2007),
  \href{http://arxiv.org/abs/0704.2296}{{\ttfamily arXiv:0704.2296 [hep-ph]}}.

\bibitem{Pagels:1979hd}
H.~{Pagels} and S.~{Stokar},
\newblock \prd {\bf 20}, 2947 (1979).

\bibitem{Roberts:1994dr}
C.~D. {Roberts} and A.~G. {Williams},
\newblock Progress in Particle and Nuclear Physics {\bf 33}, 477 (1994),
  \href{http://arxiv.org/abs/arXiv:hep-ph/9403224}{{\ttfamily
  arXiv:hep-ph/9403224}}.

\bibitem{Skullerud:2001bu}
J.~{Skullerud}, A.~{Ki Iz{\i}lers{\"u}}, and A.~G. {Williams},
\newblock Nuclear Physics B Proceedings Supplements {\bf 106}, 841 (2001).

\bibitem{Alkofer:2008tt}
R.~{Alkofer}, C.~S. {Fischer}, F.~J. {Llanes-Estrada}, and K.~{Schwenzer},
\newblock Annals of Physics {\bf 324}, 106 (2009),
  \href{http://arxiv.org/abs/0804.3042}{{\ttfamily arXiv:0804.3042 [hep-ph]}}.

\bibitem{Fischer:2009jm}
C.~S. {Fischer} and R.~{Williams},
\newblock Physical Review Letters {\bf 103}, 122001 (2009),
  \href{http://arxiv.org/abs/0905.2291}{{\ttfamily arXiv:0905.2291 [hep-ph]}}.

\bibitem{Hopfer:2013np}
M.~{Hopfer}, A.~{Windisch}, and R.~{Alkofer},
\newblock PoS Confinement X , 073 (2013),
  \href{http://arxiv.org/abs/1301.3672}{{\ttfamily arXiv:1301.3672 [hep-ph]}}.

\bibitem{Alford:2002kj}
M.~{Alford} and K.~{Rajagopal},
\newblock Journal of High Energy Physics {\bf 6}, 31 (2002),
  \href{http://arxiv.org/abs/arXiv:hep-ph/0204001}{{\ttfamily
  arXiv:hep-ph/0204001}}.

\bibitem{Nickel:2009wj}
D.~{Nickel},
\newblock \prd {\bf 80}, 074025 (2009),
  \href{http://arxiv.org/abs/0906.5295}{{\ttfamily arXiv:0906.5295 [hep-ph]}}.

\bibitem{Broniowski:2011ef}
W.~{Broniowski},
\newblock Acta Phys.Polon.Supp. {\bf 5}, 631 (2012),
  \href{http://arxiv.org/abs/1110.4063}{{\ttfamily arXiv:1110.4063 [nucl-th]}}.

\bibitem{Deryagin:1992rw}
D.~V. {Deryagin}, D.~Y. {Grigoriev}, and V.~A. {Rubakov},
\newblock International Journal of Modern Physics A {\bf 7}, 659 (1992).

\bibitem{Shuster:1999tn}
E.~{Shuster} and D.~T. {Son},
\newblock Nuclear Physics B {\bf 573}, 434 (2000),
  \href{http://arxiv.org/abs/arXiv:hep-ph/9905448}{{\ttfamily
  arXiv:hep-ph/9905448}}.

\bibitem{Bowers:2002xr}
J.~A. {Bowers} and K.~{Rajagopal},
\newblock \prd {\bf 66}, 065002 (2002),
  \href{http://arxiv.org/abs/arXiv:hep-ph/0204079}{{\ttfamily
  arXiv:hep-ph/0204079}}.

\bibitem{Casalbuoni:2003wh}
R.~{Casalbuoni} and G.~{Nardulli},
\newblock Reviews of Modern Physics {\bf 76}, 263 (2004),
  \href{http://arxiv.org/abs/arXiv:hep-ph/0305069}{{\ttfamily
  arXiv:hep-ph/0305069}}.

\bibitem{Anglani:2013gfu}
R.~{Anglani} {\em et~al.},
\newblock arXiv:1302.4264 [hep-ph]  (2013),
  \href{http://arxiv.org/abs/1302.4264}{{\ttfamily arXiv:1302.4264 [hep-ph]}}.

\bibitem{Alford:1999pa}
M.~{Alford}, J.~{Berges}, and K.~{Rajagopal},
\newblock Nuclear Physics B {\bf 558}, 219 (1999),
  \href{http://arxiv.org/abs/arXiv:hep-ph/9903502}{{\ttfamily
  arXiv:hep-ph/9903502}}.

\bibitem{Nambu:1961fr}
Y.~{Nambu} and G.~{Jona-Lasinio},
\newblock Physical Review {\bf 124}, 246 (1961).

\end{thebibliography}
\bibliographystyle{h-physrev-arxiv}

\end{document}